\documentclass[hidelinks,onefignum]{siamart220329}
\usepackage{amsfonts,amsmath,graphicx,subfigure,url}
\usepackage{amssymb}
\usepackage{algpseudocode}
\usepackage{algorithm}
\usepackage{color}
\usepackage{soul}

\newcommand{\EVH}[1]{{#1}}

\newcommand{\PAKe}[1]{{#1}}
\newcommand{\PAKr}[1]{{#1}}
\newcommand{\EVHd}[1]{{#1}}
\newcommand{\EVHe}[1]{{#1}}
\newcommand{\EVHf}[1]{{#1}}

\newcommand{\EVHR}[1]{{#1}}
\newcommand{\PBB}[1]{{#1}}

\newcommand{\PAKp}[1]{{#1}}

\newcommand{\half}{\frac{1}{2}}
\newcommand{\bu}{\bar{u}}

\newcommand{\R}{\mathbb{R}}
\newcommand{\Gz}{\Gamma_0}
\newcommand{\bmu}{\bar{\mu}}
\newcommand{\hu}{\hat{u}}
\newcommand{\dt}{\Delta t}
\newtheorem{remark}{Remark}

\title{An optimization-based coupling of reduced order models with efficient reduced adjoint basis generation approach}
\author{Elizabeth Hawkins, Paul Kuberry, Pavel Bochev}
\date{June 2024}

\begin{document}

\maketitle
\begin{abstract}

Optimization-based coupling (OBC) is an attractive alternative to traditional Lagrange multiplier  approaches in multiple modeling and simulation contexts; see, e.g., \cite{GHopt,Bochev_14_SINUM,MdBo,Abdulle_16_MMS,Gervasio_01_NM}.
However, application of OBC to time-dependent problem\PAKr{s} has been hindered by the computational cost of finding the stationary points of the associated Lagrangian, which requires primal and adjoint solves. This issue can be mitigated by using OBC in conjunction with computationally efficient reduced order models (ROM). To demonstrate the potential of this combination, in this paper we develop an optimization-based ROM-ROM coupling for a transient advection-diffusion transmission problem. \PAKr{We pursue the ``optimize-then-reduce'' path towards solving the minimization problem at each timestep and solve reduced-space adjoint system of equations, where t}he main challenge in this formulation is the generation of adjoint snapshots and reduced bases for the adjoint systems required by the optimizer. One of the main contributions of the paper is a new technique for efficient adjoint snapshot collection for gradient-based optimizers in the context of optimization-based ROM-ROM couplings. We present numerical studies demonstrating the accuracy of the approach along with comparison between various approaches for selecting a reduced order basis for the adjoint systems, including decay of snapshot energy, average iteration counts, and timings.

\end{abstract}
\section{Introduction}
\PBB{Traditional coupling methods often use Lagrange multipliers to enforce the interface compatibility conditions between the subdomains being coupled. These methods are subject to inf-sup stability requirements and can be difficult to formulate in cases such as problems with spatially non-coincident interfaces. In contrast, optimization-based couplings (OBC) minimize an objective expressing the interface conditions subject to the subdomain governing equations. Resulting PDE-constrained formulations have some attractive mathematical and computational properties, including less stringent compatibility conditions and ability to handle a wide range of interface configurations. 
These properties make OBC  a potentially attractive alternative to Lagrange multiplier-based formulations in applications ranging from domain decomposition \cite{GPHopt,Gunzburger_00_CMA,Gunzburger_99_CMA,Du_00_SINUM} to coupled multiphysics \cite{BGS,Bochev_14_SINUM,MdBo,Abdulle_16_MMS,Gervasio_01_NM,Bochev_16b_CAMWA} and non-standard interface configurations such as spatially non-coincident interfaces \cite{KBP,Bochev_18_INPROC}.}
%
%
\PBB{However, despite the flexibility afforded by OBC in handling disparate physics models and interface configurations, its utilization for time-dependent problems has been hindered by the fact that} the solution of stationary points of the Lagrangian involves primal and adjoint solves, which can be prohibitively expensive for traditional full-order models (FOMs).

Recent developments in projection-based reduced order models (ROMs), machine learned surrogates such as physics informed neural networks (PINNs), and system identification surrogate modeling techniques such as dynamic mode decomposition (DMD) \PBB{prompt a reexamination of OBC as a tool for multifidelity modeling involving couplings of reduced order models and/or full order models (FOMs).}
%

Application of \PBB{OBC} as a domain decomposition technique was introduced in \cite{GPHopt} in the context of a finite element discretization of Poisson's equation. \PBB{OBC} was extended to Navier-Stokes equations in \cite{GHopt} and later to fluid-structure interaction in \cite{KuberryLee}. Both the gradient descent algorithm and Gauss-Newton with the conjugate gradient algorithm were developed and applied in these \PAKe{works}. The gradient descent approach utilized the adjoint problem of the PDE to find a descent direction. 
\PBB{Furthermore, \cite{GHopt} showed that for the Navier-Stokes equations finite element implementations of the gradient descent and Gauss-Newton schemes attain the expected approximation accuracy.}
The methods developed in \cite{GHopt} involve solving finite element formulations of the primal and adjoint problems.

Projection-based reduced order modeling (ROM) using the Proper Orthogonal Decomposition (POD)/Galerkin method was first developed for applications to turbulence in \cite{Sirovich} and \cite{Holmes}. \PBB{POD-based ROM} has since been extended to many other applications, \PBB{including coupled problems with overlapping \cite{IoSaTa} and non\PAKe{-}overlapping interfaces \cite{Amy,Corigliano:2015}. Other relevant examples include utilization of domain decomposition to effect ROM-ROM and ROM-FOM couplings \cite{Lucia:2003,Maday_02_CRM,Maday_04_SISC}.}

\PBB{Optimization-based couplings have also been explored in the context of ROM-ROM couplings in \cite{IoSaTa}. \PAKr{That} paper introduces port and bubble bases, enabling static condensation of the bubble degrees of freedom (DOF). For elliptic PDEs the OBC formulation in this paper  is equivalent to an overlapping alternating Schwarz iteration. Although similar to the ideas in this paper, there are some important distinctions between our approach and the OBC in \cite{IoSaTa}. The latter uses the port DOF as the control, unlike our formulation which uses the interface traction force. Additionally, the OBC in \cite{IoSaTa} is developed assuming} overlapping subdomains, whereas we are targeting a non-overlapping subdomain setting.

\PBB{Adoption of existing FOM-FOM coupling techniques} for ROM-FOM and ROM-ROM couplings, \PBB{as done in many of the works cited above,} can result in unexpected difficulties that require consideration specific to the \PBB{presence} of ROMs. 
\PBB{For example, the paper  \cite{Amy} considered a Lagrange multiplier-based ROM-ROM coupling for a stationary advection-diffusion transmission problem. While the associated Ladyzhenskya-Babuska-Brezzi (LBB) stability condition can be easily satisfied for the parent FOM-FOM formulation, it's verification for the ROM-ROM problem requires an alternate construction of the reduced order basis for the Lagrange multiplier.} 

Here we make a similarly motivated investigation towards adapting an existing OBC technique \cite{GPHopt} to the ROM setting. \PAKr{Note that there are at least two choices to be made when applying ROM to OBC problems. First one can choose whether to pose the optimization problem using the reduced space system (reduce-then-optimize) or to pose the optimization problem using the full-order space and then project state equations and sensitivity equations or adjoint equations onto a reduced basis (optimize-then-reduce). Second, in computing the gradient, one can choose whether to solve many sensitivity equations (as many as modal degrees of freedom for the control) or solve an adjoint equation in the reduced space. We opt to explore the optimize-then-reduce approach with adjoint solves to form the gradient. We avoid solving multiple reduced-space sensitivity systems of equations in this way, but add the difficulty of finding an appropriate snapshot generation technique from which to produce a reduced basis for the adjoint systems.}

\PAKr{In Taddei et. al. \cite{TADDEI2024113038}, the authors focused on non-overlapping coupling via OBC as a heterogeneous domain-decomposition technique, with ROM components in the subdomains and use the reduce-then-optimize approach with a quasi-Newton optimization algorithm solving many small sensitivity equations in a reduced basis.
In Prusak et. al. \cite{PRUSAK2024}, they investigated the use of ROMs in an optimization-based coupling framework for fluid-structure interaction and proved the existence of an optimal control. They followed the optimize-then-reduce approach with reduced space adjoint solves using a reduced basis generated from snapshots of the state solutions. In contrast to these two works, we focus on an efficient technique of snapshot collection that is useful in generating a reduced basis for the adjoint system following the optimize-then-reduce approach for combining OBC with ROM, focusing particularly on advection-dominated problems.}

\PBB{Constructing a ROM for a given full order model requires collection of FOM state snapshots.
In the context of a gradient-based solution of an OBC formulation \PAKr{posed as an optimization problem with full-order state and adjoint models and then projected onto reduced bases}, this means that one will have to collect snapshots for both the state and the adjoint.} 
While the collection of \PBB{state} snapshots to produce a reduced basis for the \PBB{state equation} has been well researched, the collection of snapshots for producing a reduced basis for the adjoint \PBB{equation} is less so. The common method of snapshot collection for the adjoint solves the entire FOM-FOM coupled problem using gradient descent and stores all calculated adjoints as snapshots. \PAKe{There are several disadvantages to that approach. First, it requires sequentially solving the state, adjoint, and then control update iteratively (variable number of iterations per time step) and solutions at a particular timestep rely on the quality of solutions of all prior timesteps.}

The snapshot collection technique we present \PAKr{here} in this \PAKr{paper} breaks the connection between adjoint snapshots and previous solutions and allows for control over a fixed number of snapshots per timestep. \PAKr{We also here only investigate using our approach to generate snapshots of adjoint solutions for a reduced basis for the adjoint system. Considering the relevance of our investigation to the certified basis approach \cite{Negri2013,Grepl2014,Bader2016}, where full order state and adjoint solutions play a critical role in generating a posteriori error estimates for linear-quadratic optimal control problems, this could be considered a practical approach for generating the adjoint snapshots and accompanying aggregate reduced space. We limit ourselves here in this work to considering state snapshot generated reduced bases for state systems and either state or adjoint snapshot generated reduced bases for the adjoint system, but in future work we will also investigate aggregating these reduced bases for projection of both systems as is done in that community.} 

\PBB{Following \cite{GPHopt}, in this paper} we use the gradient descent algorithm to solve the PDE-constrained optimization problem \PBB{resulting from an OBC formulation of a coupled} advection-diffusion equation. We then adapt the method to a projection-based ROM setting in order to improve on the computational time in each optimization iteration. 
\PBB{To that end,} 
we use \PAKe{a technique we term \emph{the modified gradient descent for the reduced adjoint} (MGD$m$RA)}, where \PBB{$m$} is the number of iterations stored per timestep, to improve the memory storage and computational efficiency of the offline stage without hindering the methods performance in the online stage. 

\PAKe{The model problem and coupling approach are described in Section \ref{sec:opt-problem-formulation}, gradient descent solution technique for FOM and ROM settings is described in Section \ref{sec:opt-methods}, and the MGD$m$RA snapshot collection technique is presented in Section \ref{sec:mgdnra}. Numerical results are provided in Section \ref{sec:numericalresults}, comparing the optimization iterations and accuracy of the ROM-ROM coupling approach using various snapshot collection techniques, and also comparing these to the optimization-based FOM-FOM coupling approach.}

\section{Model problem and coupling formulation}\label{sec:opt-problem-formulation}

\begin{figure}[H]
    \centering
    \includegraphics[scale=0.35,trim={6cm 8cm 6cm 5cm},clip]{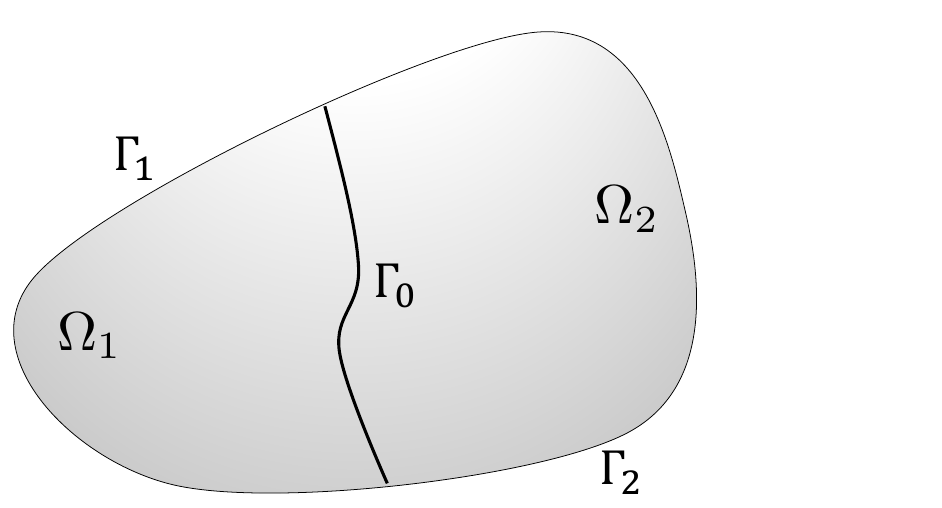}
    \caption{Example of two non-overlapping domains sharing an interface}
    \label{fig:fig2domains}
\end{figure}

\PBB{We consider an open bounded region $\Omega\subset \mathbb{R}^2$ divided into two non-overlapping subdomains $\Omega_1$ and $\Omega_2$ by an interface $\Gz$.}
We define $\Gamma_1=\partial\Omega_1\setminus\Gz$ and similarly $\Gamma_2=\partial\Omega_2\setminus\Gz$. 
\PBB{On each subdomain $\Omega_i$ we consider an advection-diffusion equation given by}
$$\begin{cases}
    u_{i,t}-\nabla\cdot\sigma_i(u_i)=f_i&\text{ in } \Omega_i\\
    u_i=0 & \text{ on } \Gamma_i
\end{cases}$$
where \PBB{$u_i$ is the concentration, $\nu_i$ is the diffusivity constant, $\mathbf{a}_i$ is the advection field, and $\sigma_i(u)= \nu_i\nabla u- \mathbf{a}_i u$ is the flux}. We couple \PBB{the subdomain} equations using \PBB{standard interface compatibility conditions,  expressing continuity of the states and the fluxes on $\Gz$:}
\begin{equation}\label{eqn:ad-fully-coupled-problem}
\begin{cases}
    u_{i,t}-\nabla\cdot\sigma_i(_i)=f_i&\text{ in } \Omega_i\\
    u_i=0 & \text{ on } \Gamma_i\\
    u_1=u_2 & \text{ on } \Gz\\
    \sigma_1(u_1)\cdot n_1 = - \sigma_2(u_2)\cdot n_2 & \text{ on } \Gz
\end{cases}
\end{equation}
where $n_i$ is the outward unit normal vector on $\Gamma_i$. 
\PBB{The coupled problem \eqref{eqn:ad-fully-coupled-problem} is representative of the type of couplings that are ubiquitous in} multi-physics problems, domain decomposition problems, and transmission problems. \PBB{In this paper we focus on optimization-based coupling formulations for this problem in which one or both subdomain equations are approximated by a reduced order model.}

\PBB{To develop the OBC formulation for \eqref{eqn:ad-fully-coupled-problem}, we shall decouple this system} using a control. While there are many choices on how to introduce a control, we follow the \PBB{approach in} \cite{GPHopt} and introduce an unknown function $g\in L^2(\Gz)$ that satisfies
\begin{equation}
\label{eqn:strong-form-ad-with-g}\begin{cases}
    u_{i,t}-\nabla\cdot\sigma_i(u_i)=f_i&\text{ in } \Omega_i\\
    u_i=0 & \text{ on } \Gamma_i\\
    \sigma_i(u_i)\cdot n_i = (-1)^i g & \text{ on } \Gz.
\end{cases}
\end{equation}

Note that the \PBB{state continuity condition} $u_1=u_2 \text{ on } \Gz$ \PAKe{is not included in \eqref{eqn:strong-form-ad-with-g}}. This is enforced later in \PAKe{an} optimization problem by the introduction of the measure of violation of this constraint in the objective function. 

 \PBB{Let} $(\cdot,\cdot)$ be the $L^2$ inner product on $\Omega_i$ for $i=1,2$ and let $(u,v)_{\Gamma_i}$ be the $L^2$ inner product on $\Gamma_i$ for $i=0,1,2$. 
 \PBB{We then discretize in time the standard weak formulation of \eqref{eqn:strong-form-ad-with-g} by using backward Euler to obtain the following semi-discrete in time problem}.
\begin{multline}\label{eqn:WF}
    \text{Find } u_i^n\in \EVHf{V}_i \text{ satisfying }\\
    \frac{1}{\dt}(u^n_i-u^{n-1}_i,v)+(\sigma_i(u^n_i),\nabla v)=(f_i^n,v)
    +(-1)^{i}(g^n,v)_{\Gz} \hspace{5pt}\forall v\in V_i,
\end{multline}
where $V_i=\{v\in H^1(\Omega_i): v=0\text{ on }\Gamma_i\}$, $g^n:=g(x,t^n)$, and $f_i^n:=f_i(x,t^n)$ for $i=1,2$. 



We \PBB{then} define the functional
\begin{equation}\label{eqn:functional}
    J_\delta(u^n_1,u^n_2,g^n):=\half \|u^n_1-u^n_2\|_{\Gz}^2 + \half \delta \|g^n\|_{\Gz}^2.
\end{equation}
This functional enforces the interface \PBB{coupling} condition $u_1=u_2 \text{ on } \Gz$ at time $t^n$ by minimizing \PBB{the state mismatch}  $ \|u^n_1-u^n_2\|_{\Gz}^2$ and also includes regularization of the optimization problem by minimizing the $L^2$ norm of $g^n$. \PAKe{Letting $\delta=0$, it is clear that a choice of $g=\sigma_2(u_2)\cdot n_2$ is a global minimizer to the minimization problem \eqref{eqn:min} and is a solution to the fully-coupled problem \eqref{eqn:ad-fully-coupled-problem}}. Furthermore, if $g$ is known, then both of the subdomain problems are independent and consequently can be solved simultaneously. The purpose of the stabilization term is to assist convergence of the gradient descent approach. Additionally, this term is of value \PAKe{for regularizing ill-posed minimization problems such as when considering} nonlinear governing equations, although they are not considered in this paper. In the case of $\delta=0$, an optimal solution will only enforce $u_1=u_2$ weakly$\text{ on }\Gz$. However, for $\delta>0$, the second term competes with the first, ultimately limiting how well we can enforce $u_1=u_2$. This balance between terms should be considered when choosing a value for $\delta$.

\PBB{\EVHR{\PAKr{For} each time step, $t^n$}, an optimization-based coupling formulation for \eqref{eqn:strong-form-ad-with-g} is given by the following constrained minimization problem}
\begin{equation}\label{eqn:min}
\begin{aligned}
     &\min J_\delta (u^n_1,u^n_2,g^n)\\
     &(u^n_1,u^n_2,g^n)\in \EVHf{V}_1\times \EVHf{V}_2\times L^2(\Gz)\\
   & \text{s.t } \eqref{eqn:WF}.
\end{aligned}
\end{equation}
To \PBB{solve \eqref{eqn:min}}, we find the \PBB{first-order} optimality conditions \PBB{defined by setting the first variation of the Lagrangian associated with \eqref{eqn:min} to zero.} The Lagrangian is given by
\begin{multline*}
    \mathcal{L}(u^n_1,u^n_2,g^n,\mu_1,\mu_2):=J_\delta (u^n_1,u^n_2,g^n)+ \sum_{i=1}^2[\frac{1}{\dt}(u^n_i-u^{n-1}_i,\mu_i)+(\sigma_i(u_i^n),\nabla \mu_i)\\
    -(f_i^n,\mu_i)+(-1)^{i+1}(g^n,\mu_i)_{\Gz}]
\end{multline*}
where $\mu_i\in V_i$ is the adjoint variable. We take the gradient of $\mathcal{L}$ with respect to $u^n_i$ and set it equal to zero to get the adjoint equation
\begin{equation}\label{eqn:adjoint}
\frac{1}{\dt}(\mu_i,\eta)+(\nu_i\nabla \mu_i + \mathbf{a}_i \mu_i,\nabla\eta)=(-1)^{i}(u^n_1-u^n_2,\eta)_{\Gz}\hspace{3pt} \forall\eta\in V_i.
\end{equation}
Similarly, the gradient of $\mathcal{L}$ with respect to $g^n$ set equal to zero yields \PBB{the equation}
\begin{equation}\label{eqn:partialg}
    \delta (\psi,g^n)_{\Gz}= -(\psi,\mu_1 -\mu_2)_{\Gz}\hspace{3pt} \forall \psi\in L^2(\Gz).
\end{equation}
Note that the gradient of $\mathcal{L}$ with respect to $\mu_i$ will return \eqref{eqn:WF}. Equations \eqref{eqn:adjoint}, \eqref{eqn:partialg}, and \eqref{eqn:WF} are the optimality conditions that a global minimizer must satisfy.\\



\section{\PBB{A gradient descent solution of the OBC problem}}\label{sec:opt-methods}
It is clear that we can \PBB{view} $u_i^n$ as a function of $g^n$ because $g^n$ provides a boundary condition of the boundary value problem for $u_i^n$. \PBB{This interpretation transforms $J_\delta$ into a function of $g^n$, which we denote as $\mathcal{M}_\delta$, i.e.,}
\begin{equation*}
    \mathcal{M}_\delta(g^n):= J_\delta(u^n_1(g^n),u^n_2(g^n),g^n)
\end{equation*}
where $u_i^n(g^n)$ is the solution to \eqref{eqn:WF} for the given $g^n$. The gradient of $\mathcal{M}_\delta(g^n)$ with respect to variation in the control, $\tilde{g}^n$, is given by
\begin{equation*}
    \frac{d \mathcal{M}_\delta(g^n)}{d g^n}\cdot \tilde{g}^n = \delta(g^n,\tilde{g}^n)_{\Gz}+(u^n_1-u^n_2,\tilde{u}^n_1-\tilde{u}^n_2)_{\Gz} \hspace{5pt}\forall \tilde{g}^n\in L^2(\Gz)\,,
\end{equation*}
where $\tilde{u}_i^n$ is the solution to the sensitivity equation 
\begin{equation}\label{eqn:sensitivity}
    \frac{1}{\dt}(\tilde{u}_i^n, v)+(\sigma_i(\tilde{u}_i^n),\nabla v)=(-1)^{i}(\tilde{g}^n,v)_{\Gz}\hspace{5pt}\forall v\in V_i.
\end{equation}


In \eqref{eqn:sensitivity} let $v=\mu_i$ and in \eqref{eqn:adjoint} let $\eta=\tilde{u}^n_i$. Combining these we get
\begin{equation}\label{eqn:intermediate-gtilde}
    (\tilde{g}^n, \mu_1-\mu_2)_{\Gz}=(\tilde{u}_1^n-\tilde{u}_2^n,u_1^n-u_2^n)_{\Gz}.
\end{equation}
Using \eqref{eqn:intermediate-gtilde}, we get that the gradient of the functional is 
\begin{equation}\label{eq:grad}
    \frac{d \mathcal{M}_\delta(g^n)}{d g^n}\cdot \tilde{g}^n = \delta(g^n,\tilde{g}^n)_{\Gz}+(\tilde{g}^n, \mu_1-\mu_2)_{\Gz} \hspace{5pt}\forall \tilde{g}^n\in L^2(\Gz).
\end{equation}
Because $\tilde{g}^n$ is an arbitrary function, it must hold that 
$$\frac{d \mathcal{M}_\delta(g^n)}{d g^n} = \delta g^n+ (\mu_1-\mu_2)|_{\Gz}.$$
This allows us to use the gradient descent method where the $k^{th}$ iteration is defined by
\begin{equation*}
    g^{n,(k)}=g^{n,(k-1)} -\alpha\frac{d \mathcal{M}_\delta(g^n)}{d g^n}.
\end{equation*}
\PBB{Inserting \eqref{eq:grad} into the above formula then yields the following expression for the update:}
\begin{equation*}
    g^{n,(k)}=(1-\alpha\delta)g^{n,(k-1)} -\alpha(\mu^{(k)}_1-\mu^{(k)}_2)|_{\Gz}\,,
\end{equation*}
where $\alpha$ is \PBB{a relaxation parameter} and $\mu_i^{(k)}$ is the solution to the adjoint equation for $g^{n,(k-1)}$. This method follows the approach in \cite{GPHopt} which has been generalized in \cite{Gunzbook}. 

For any iterative optimization algorithm, a stopping criteria is required. We halt the optimization algorithm when 
$$\EVHf{J_\delta(u_1^n,u_2^n,g^{n,(k-1)})}=\EVHf{\frac{1}{2}\|u_1^n-u_2^n\|_{\Gz}^2 +\frac{\delta}{2}\|g^{n,(k-1)}\|^2_{\Gz}}< tol
$$ 
where $u_i^n$ is the state equation solution on $\Omega_i$ for a particular $g^{n,(k-1)}$. Recall that the purpose of the functional $J_\delta$ is to minimize $\|u_1^n-u_2^n\|_{\Gz}$. The \PBB{value of} $\alpha$ at each optimization iteration can be chosen by the user, either empirically or through a more sophisticated approach such as the use of a line search.

\begin{algorithm}[H]
\caption{Continuous form of the gradient descent method}\label{alg:GDCont}
\begin{algorithmic}
    \State Given $g^{n,(0)}$ 
    \For{$k=1,2,...$}\\
    \State Solve \eqref{eqn:WF} using $g^{n,(k-1)}$ for $u^n_1$ and $u^n_2$\\
    \State Solve \eqref{eqn:adjoint} using $u^n_1$ and $u^n_2$ for $\mu_1^{(k)}$ and $\mu_2^{(k)}$\\
    \State $g^{n,(k)}= (1-\alpha\delta)g^{n,(k-1)} -\alpha(\mu^{(k)}_1-\mu^{(k)}_2)|_{\Gz}$\\
    \EndFor
\end{algorithmic}
\end{algorithm}

\subsection{Application of finite element models to \PBB{OBC}}\label{FEMsection}
The symbol $\Omega_{i,h}$ will stand for a conforming quasi-uniform partition of $\Omega_i$ into finite elements $\{\phi_{i,s}\}_{s=1}^{N_i}$, and mesh parameter $h_i$, where $N_i=|\{\phi_{i,s}\}|$ is the cardinality of the finite dimensional basis. \EVH{Let $V^h_i=\{u\in H^1(\Omega_{i,h}): u=0 \text{ on }\Gamma_i\}$}. \PAKe{We define the induced interface finite element space $G_i^h$ as the trace of the interface part of $V_i^h$, i.e., $G_i^h$ = $V_i^h|_{\Gamma_0}$ and $\xi_{i,j}=\phi_{i,j}|_{\Gamma_0}$, where $\{\xi_{i,j}\}$ is the set of all basis functions whose restriction to $\Gamma_0$ is not exactly zero everywhere. The basis for $G_i^h$ consists of finite elements $\xi_{i,j}$, $j\in\{1,...,|G_i^h|\}$. While it is possible to discretize the control $g^n$ on $\Gamma_0$ with some basis $G_{\Gamma_0}^h$ with basis elements $\xi_{\Gamma_0}$ independently of $\xi_{i}$, we define $G_{\Gamma_0}^h=G_1^h$ and $\xi_{\Gamma_0}=\xi_{1}$ in this work.}
\PAKr{Note that for subdomains with matched vertices on the interface, $G_1^h=G_2^h$. This is the case that we will consider for numerical experiments in this paper.}

 We discretize in space \EVHd{using $\phi_{i,k}$} to get the Galerkin formulation of \eqref{eqn:WF}
\begin{multline}\label{eqn:GWF}
    \text{Find } u^{h,n}_i\in V^h_i \text{ satisfying }\\
    \frac{1}{\dt}(u^{h,n}_i-u^{h,n-1}_i,v^h)+(\sigma_i(u^{h,n}_i),\nabla v^h)=(f_i^n,v^h)+(-1)^{i}(g^n,v^h)_{\Gz},\hspace{5pt} \forall v^h\in V^h_i.
\end{multline}
This is equivalent to the following matrix system:
\begin{multline}\label{eqn:FEMwfmatrix}
\text{Find } \bu_i^n\in \R^{N_i} \text{ satisfying }\\
    \frac{1}{\dt}M_i  \bu_{i}^n +(\nu_i K_i - A_i)\bu_i^n = \bar{f}^n_i+(-1)^i M_{\Gz,i}\bar{g}^n+\frac{1}{\dt}M_i \bu_i^{n-1}\,,
\end{multline}
where $(M_i)_{k,j}:= (\phi_{i,k},\phi_{i,j})$, $(K_i)_{k,j}:= (\nabla \phi_{i,k},\nabla \phi_{i,j})$, $(A_i)_{k,j}:= ({\bf{a}} \phi_{i,k},\nabla \phi_{i,j})$, $(\bar{f}^n_i)_k:=(f_i^n,\phi_{i,k})$, and $(M_{\Gamma_0,i})_{k,j}:= (\xi_{\Gamma_0,j}, \xi_{i,k})_{\Gz}$. \PAKr{We also note that $(M_{\Gamma_0,\Gamma_0})_{k,j}:= (\xi_{\Gamma_0,j}, \xi_{\Gamma_0,k})_{\Gz}$, which will be used later.}  

One popular way of dealing with Dirichlet boundary conditions in finite elements is to eliminate known degrees of freedom from the rows and columns of the matrix, adjusting the right hand side accordingly. The $M_i$'s above do not reflect this modification, and this is only mentioned here as it becomes relevant for their reference in the ROM-specific section that follows.

 We will also need to solve the adjoint equation \eqref{eqn:adjoint} 
 using FEM. So we discretize in space using the same basis as \eqref{eqn:GWF} to get the discrete adjoint equation
\begin{equation}\label{eqn:FEMadjoint}
    \frac{1}{\dt}(\mu^h_i,\eta^h)+(\nu_i\nabla \mu_i^h + \mathbf{a}_i \mu_i^h,\nabla \eta^h)=(-1)^{i}(u^{h,n}_1-u^{h,n}_2,\eta^h)_{\Gz},\hspace{5pt} \forall\eta^h\in V^h_i.
\end{equation}
The discrete adjoint equation is equivalent to the matrix equation
\begin{equation}\label{eqn:FEMadjointmatrix}
    \frac{1}{\dt}M_i \bmu_i+(\nu_i K_i+A_i^T)\bmu_i = (-1)^i M_{\Gz,i}(I_{1\rightarrow 0}\bu_1-I_{2\rightarrow 0}\bu_2).
\end{equation}


\PAKe{We discretize the continuous gradient descent algorithm, Algorithm \ref{alg:GDCont}, and provide it below in Algorithm \ref{alg:GDFEM}. We solve \eqref{eqn:FEMwfmatrix} and \eqref{eqn:FEMadjointmatrix} as discretized instances of \eqref{eqn:WF} and \eqref{eqn:adjoint}. Then, the FEM gradient descent algorithm solves for updates to the coefficient vector $\bar{g}^n$ and uses the vector solutions to \eqref{eqn:GWF} and \eqref{eqn:FEMadjoint} to update $\bar{g}^n$ at each iteration. The algorithm that follows uses $I_{i\rightarrow 0}$ to denote an interpolation operator from $V^h_i$ to the space of the control $g^n$, $G_{\Gamma_0}$, for i=1,2.}

\begin{algorithm}[H]
\caption{Finite element form of the gradient descent method}\label{alg:GDFEM}
\begin{algorithmic}
    \State Given $\bar{g}^{n,(0)}$ 
    \For{$k=1,2,...$}\\
    \State Solve \eqref{eqn:FEMwfmatrix} using $\bar{g}^{n,(k-1)}$ for $\bu^{n}_1$ and $\bu^{n}_2$\\
    \State Solve \eqref{eqn:FEMadjointmatrix} using $\bu^{n}_1$ and $\bu^{n}_2$ for $\bmu^{(k)}_1$ and $\bmu^{(k)}_2$\\
    \State $\bar{g}^{n,(k)}= (1-\alpha\delta)\bar{g}^{n,(k-1)} -\alpha(I_{1\rightarrow 0}\bmu^{(k)}_1-I_{2\rightarrow 0}\bmu^{(k)}_2)|_{\Gz}$\\
    \EndFor
\end{algorithmic}
\end{algorithm}

\begin{remark}
The scope of this paper only includes investigating matched meshes at the interface, with $I_{1\rightarrow 0}$ as the identity operator and $I_{2\rightarrow 0}$ as a permutation matrix that matches degrees of freedom from $\EVHf{V}_2^{h}$ on the interface consistent with DOF ordering for $\Omega_1$ (consistent with the definition of $\bar{g}^n$).
\end{remark}

\begin{remark}
Applying the Galerkin finite element method to \eqref{eqn:WF} on moderately refined meshes can produce strongly oscillatory solutions. We stabilize using the Streamline Upwind Petrov-Galerkin (SUPG) stabilization approach \cite{BrHu}. This introduces an option to either include the SUPG stabilization terms in the state equations and then derive the adjoint equation (discretize-then-optimize), or to formulate the optimization problem as in \eqref{eqn:min} and then add SUPG stabilization to \eqref{eqn:WF} and \eqref{eqn:adjoint}, consistent with the respective subproblem (optimize-then-discretize). We choose to optimize-then-discretize later in this work, which is similar in accuracy for linear finite elements and more accurate for higher order finite elements when compared with the discretize-then-optimize approach. A thorough investigation contrasting the discretize-then-optimize and optimize-then-discretize approaches, applied to the advection-diffusion equations, can be found in \cite{CoHe}.
\end{remark}

\subsection{Application to \PBB{OBC} of projection-based ROMs}\label{ROMsection}

Before describing \PBB{extension} of the \PBB{OBC} approach from FOM-FOM to ROM-ROM coupling, we will briefly describe \PBB{the construction of the} projection-based reduced order models \PBB{used in this work}. The generation of our ROM basis closely follows the detailed description in \cite{deCKu}. \PAKr{We solve a monolithic, uncoupled formulation of the problem over $\Omega = \Omega_1 \cup \Omega_2$ with a uniform time step and store snapshots of the solution. We then decompose the single-domain ($\Omega$) solution, segregating the DOFs from the monolithic solution into two snapshot matrices, each snapshot matrix containing portions of the solution relevant to its respective subdomain}. We note that there are other ways to generate the snapshot matrices corresponding to each subdomain, \PBB{but using a monolithic FEM solution ensures the ``best possible'', in terms of accuracy, snapshots}. Similarly to \cite{deCKu}, we remove contributions to the snapshot columns for DOFs corresponding to Dirichlet nodes. A singular value decomposition of each of the snapshot matrices, altered as described and related to boundary conditions, is performed. 
While a threshold criteria is often used to determine how many singular values and left singular vectors are retained, in this work we choose how many singular vectors to keep independent of a threshold criteria.

Let $\{\psi_{i,k}\}_{k=1}^{N_{i,r}}$ be a ROM basis associated with the solution to \eqref{eqn:GWF} where $N_{i,r}<< N_i$ and $N_i$ is the size of the FEM discretized system in $\Omega_i$. 
\PAKr{Given a reduced basis $\{\psi_{i,k}\}_{k=1}^{N_{i,r}}$, the next step is to project the FOM onto this basis. Here, we restrict
our discussion to an approach called “discrete Galerkin projection”, where the governing equations in their discretized form are projected onto the POD basis in the discrete \EVHR{$l^2$} inner product.} To handle inhomogeneous  Dirichlet boundary conditions $u_i=\beta$ on $\Gamma_i$, we follow the approach in \cite{GuPeSh} and write the change to a reduced basis representation as 
\begin{equation}\label{eqn:primalROMrep}
    \EVHd{\Psi_{u,i}} \hu_i^n+\bar{\beta}_i^n=\bu_i^n;\quad \hu_i\in\R^{N_{i,r}}
\end{equation}
\PBB{where $\hu_i$ is the coefficient vector of the reduced state on $\Omega_i$ and} $\EVHd{\Psi_{u,i}}\in \R^{N_i \times N_{i,r}}$ is the matrix whose columns are the ROM basis elements \PAKe{formed from snapshots of the state equations solutions} and $\bar{\beta}_i^n \in \R^{N_i}$ is a column vector such that 
$$(\bar{\beta}^n_i)_k:=\begin{cases}
    0 & \text{ if } x_k \text{ is not a boundary node on } \Gamma_i\\
    \beta_i(x_k,t^n) & \text{ if } x_k \text{ is a boundary node on } \Gamma_i
\end{cases}.$$
Note that this change of basis accounts for the inhomogeneous Dirichlet boundary conditions, \PAKe{which means that the} matrix equations do not need to be modified in the way previously described for FEM. This method of boundary condition enforcement for ROM is described in greater detail as Method 1 in \cite{GuPeSh}.

We use \PBB{the ansatz \eqref{eqn:primalROMrep}} for $\bar{u}_i^n$ and left multiply by $\EVHd{\Psi_{u,i}}^T$ in \eqref{eqn:FEMwfmatrix} to get the ROM matrix equation for the Galerkin weak form.

Find $\hu_i^n\in \R^{N_{i,r}}$ satisfying
\begin{equation}\label{eqn:ROMwfmatrix}
    \frac{1}{\dt}\hat{M}_i \hu_i^n+(\nu_i \hat{K}_i - \hat{A}_i)\hu_i^n =\\ \hat{f}_i^n+ (-1)^i M_{\Gz,i}\bar{g}+\frac{1}{\dt}\hat{M}_i \hu_i^{n-1}, 
\end{equation}
where $\hat{M}_i :=\EVHd{\Psi_{u,i}}^T M_i \EVHd{\Psi_{u,i}}$, $\hat{K}_i=\EVHd{\Psi_{u,i}}^T K_i\EVHd{\Psi_{u,i}}$, $\hat{A}_i:=\EVHd{\Psi_{u,i}}^T A_i\EVHd{\Psi_{u,i}}$,  and $\hat{f}_i^n:=\EVHd{\Psi_{u,i}}^T\bar{f}_i^n-\EVHR{\frac{1}{\dt}}\EVHd{\Psi_{u,i}}^T M_i (\bar{\beta}_{i}^n-\bar{\beta}_{i}^{n-1})-\EVHd{\Psi_{u,i}}^T(\nu_i K_i\EVHR{-}A_i)\bar{\beta}_i^n$.

We also use the change of basis operation on $\bmu_i$ given by
\begin{equation*}
    \EVHd{\Psi_{\mu,i}} \hat{\mu}_i=\bar{\mu}_i \text{ for some } \hat{\mu}_i\in\R^{N_{i,r}},
\end{equation*}
\PAKe{where $\Psi_{\mu,i}$ is a reduced basis generated from snapshots of the adjoint equation}, and apply the same process to \eqref{eqn:FEMadjointmatrix} to get the  adjoint ROM matrix equation
\begin{multline}\label{eqn:ROMadjointmatrix}
    \frac{1}{\dt}\hat{M}_i \hat{\mu}_i+(\nu_i \hat{K}_i+\hat{A}_i^T)\hat{\mu}_i = (-1)^i(\EVHd{\Psi_{\mu,i}}^TM_{\Gz,i} (I_{1\rightarrow 0}\Psi_{u,1}\hu_1^n-I_{2\rightarrow 0}\Psi_{u,2}\hu_2^n)\\
    +\EVHd{\Psi_{\EVHR{\mu},i}}^T\EVHR{M_{\Gz,i}}(I_{1\rightarrow 0}\bar{\beta}_1^n-I_{2\rightarrow 0}\bar{\beta}_2^n)).
\end{multline}
where $\hat{M}_i :=\EVHd{\Psi_{\mu,i}}^T M_i \EVHd{\Psi_{\mu,i}}$, $\hat{K}_i=\EVHd{\Psi_{\mu,i}}^T K_i\EVHd{\Psi_{\mu,i}}$, and $\hat{A}_i:=\EVHd{\Psi_{\mu,i}}^T A_i\EVHd{\Psi_{\EVHR{\mu},i}}$.


We solve the finite element gradient descent method projected onto a reduced basis, i.e.  \eqref{eqn:FEMwfmatrix} and \eqref{eqn:FEMadjointmatrix} in Algorithm \ref{alg:GDFEM} becomes \eqref{eqn:ROMwfmatrix} and \eqref{eqn:ROMadjointmatrix} in Algorithm \ref{alg:GDROM}. However, because we've chosen to keep $\bar{g}^n$ in the full order space we must project the reduced space solutions to the adjoint problems into the full order space in order to subtract them from $\bar{g}^n$. This projection is simply $\EVHf{\bmu_i}=\EVHd{\Psi_{\mu,i}} \EVHf{\hat{\mu}_i}$. Again, in the algorithm that follows, we use $I_{i\rightarrow 0}$ to denote an interpolation operator from $V^h_i$ to the space of the control $\bar{g}^n$, for i=1,2.\\

\begin{algorithm}[H]
\caption{ROM form of the gradient descent method}\label{alg:GDROM}
\begin{algorithmic}
    \State Given $\bar{g}^{n,(0)}$ 
    \For{$k=1,2,...$}\\
    \State Solve \eqref{eqn:ROMwfmatrix} using $\bar{g}^{n,(k-1)}$ for $\hat{u}^{n}_1$ and $\hat{u}^{n}_2$\\
    \State Solve \eqref{eqn:ROMadjointmatrix} using $\hat{u}^{n}_1$, and $\hat{u}^{n}_2$ for $\hat{\mu}^{(k)}_1$ and $\hat{\mu}^{(k)}_2$\\
    \State $\bar{g}^{n,(k)}= (1-\alpha\delta)\bar{g}^{n,(k-1)} -\alpha(I_{1\rightarrow 0}\Psi_{\mu,1}\hat{\mu}^{(k)}_1-I_{2\rightarrow 0}\Psi_{\mu,2}\hat{\mu}^{(k)}_2)|_{\Gz}$\\
    \EndFor
\end{algorithmic}
\end{algorithm}

\PAKr{To expose the structure of the discretized optimization problem, we include here the optimality system combining \eqref{eqn:ROMwfmatrix}, \eqref{eqn:ROMadjointmatrix}, and a discretized version of \eqref{eqn:partialg}.}


\begin{equation}\label{eqn:monolithic-optimality-system-reduced}
\begin{bmatrix}
    B_1 & 0 & B_2\\
    B_3 &B_4&0\\
    0& B_5& \delta M_{\Gz,\Gz}
\end{bmatrix}
\begin{bmatrix}
  \hat{u}_1\\
  \hat{u}_2\\
  \hat{\mu}_1\\
  \hat{\mu}_2\\
  g
\end{bmatrix}
=
\begin{bmatrix}
   \hat{f}_1^n+ \frac{1}{\dt}\hat{M}_1 \hu_1^{n-1}\\
   \hat{f}_2^n+ \frac{1}{\dt}\hat{M}_2 \hu_2^{n-1}\\
   \Psi_{\mu,1}^TM_{\Gz,1}(I_{1\rightarrow 0}\bar{\beta}_1^n-I_{2\rightarrow 0}\bar{\beta}_2^n))\\
   \Psi_{\mu,2}^TM_{\Gz,2}(I_{1\rightarrow 0}\bar{\beta}_1^n-I_{2\rightarrow 0}\bar{\beta}_2^n))\\
   0
\end{bmatrix}
\end{equation}
where 
\begin{align*}
B_1&=\begin{bmatrix}
\frac{1}{\dt}\hat{M}_1 +(\nu_1 \hat{K}_1 - \hat{A}_1) & 0\\
0& \frac{1}{\dt}\hat{M}_2+(\nu_2 \hat{K}_2 - \hat{A}_2)
\end{bmatrix},\\ 
B_2&=\begin{bmatrix}
\Psi_{u,1}^T M_{\Gz,1}\\
- \Psi_{u,2}^T M_{\Gz,2}
\end{bmatrix},\\ 
B_3&=\begin{bmatrix}
\Psi_{\mu,1}^TM_{\Gz,1}I_{1\rightarrow 0}\Psi_{u,1}& -\Psi_{\mu,1}^TM_{\Gz,1}I_{2\rightarrow 0}\Psi_{u,2} \\
-\Psi_{\mu,2}^TM_{\Gz,2}I_{1\rightarrow 0}\Psi_{u,1}&\Psi_{\mu,2}^TM_{\Gz,2}I_{2\rightarrow 0}\Psi_{u,2}
\end{bmatrix},\\ 
B_4&=\begin{bmatrix}
\frac{1}{\dt}\hat{M}_1+(\nu_1 \hat{K}_1+\hat{A}_1^T)  &0 \\
0 &\frac{1}{\dt}\hat{M}_2+(\nu_2 \hat{K}_2+\hat{A}_2^T)
\end{bmatrix},\\ 
B_5&=\begin{bmatrix}
M_{\Gz,\Gz}I_{1\rightarrow 0}\Psi_{\mu,1}& -M_{\Gz,\Gz}I_{2\rightarrow 0}\Psi_{\mu,2}
\end{bmatrix}.
\end{align*}

\PAKr{With the application of projection to a reduced basis for the state and adjoint systems, this monolithic optimality system \eqref{eqn:monolithic-optimality-system-reduced} is reasonable to construct and solve with, rather than using an iterative optimization scheme, as the linear system size is greatly reduced through projection onto reduced bases. We will investigate using the monolithic linear system for the optimality system as a forward solver in future work, although we note that an iterative optimization approach will still be required to obtain snapshots from which to produce a reduced basis for the adjoint system and the snapshot collection technique that we are about to introduce will remain applicable.}

\section{Snapshot collection methods}\label{sec:mgdnra}

\PBB{Since the main focus of this work is on the formulation and efficient solution of an optimization-based coupling (OBC) of reduced order models, we shall not discuss in detail the generation of the ROMs for the subdomain equations. Thus, we shall adopt the viewpoint that projection-based subdomain  ROMs are already available and our task is to implement their coupling through a PDE-constrained optimization approach. From this vantage point} \PAKe{generation of the reduced bases for these ROMs assumes snapshots of the state equation solutions are already available. We will make use of this assumption in our new snapshot collection technique that is the main topic of this section.}

\PBB{Broadly speaking there are two distinct ways to approach the construction of a ROM basis for the adjoint equation. The first one is to simply reuse the  state snapshots, which we assume are readily available.} This method is appealing because \PBB{it does not} require \PBB{any additional snapshot collection procedure. However, it is not guaranteed to provide a good reduced basis for the adjoint equation because the adjoint states are fundamentally different from the solutions of our equations.}

\PBB{A better approach would be to collect separate snapshots for the adjoint from \eqref{eqn:adjoint} and use these ``custom'' snapshots to define a reduced basis for the adjoint equation.} However, this  requires solving the FOM-FOM coupled problem using gradient descent at every time step, which \PAKe{may be} impractical. This is necessary for two reasons. First, \PBB{one needs} access to \PBB{the quantity} $u_1^{h,n}-u_2^{h,n}$ on the interface such that  $u_1^{h,n}-u_2^{h,n}|_{\Gz} \neq 0$. This \PBB{quantity} \PAKe{will be} non-zero when \eqref{eqn:GWF} is solved using an non-optimal \PAKe{choice of the control} $g^n$. Second, \eqref{eqn:GWF} depends on the solution to the previous time step $u_i^{h,n-1}$, therefore we have to solve the entire \PAKe{optimization problem \PBB{while using a} tight convergence tolerance at each time step. The tolerance must be tight, not necessarily because snapshots of the adjoint must be very accurate, but rather because the optimization problem at the next time step uses the state solution from the previous time step.}


\PBB{In this paper we formulate an efficient alternative to the latter procedure, which} solves a modified optimization problem to collect snapshots of adjoint solutions. 
\PBB{We term this approach \emph{Modified Gradient Descent for the  Reduced Adjoint}} (MGD$m$RA) basis. The modified optimization problem will use the state snapshots $U_{snap}$ to make \eqref{eqn:GWF} independent of the previous time steps, meaning it is independent of the computed solutions of the previous time steps.
\PAKe{
\begin{multline}\label{eqn:ModWF}
    \text{At timestep $n$, for some iteration $k\ge1$, find } u_i^{h,(k)}\in \EVHf{V}^{h}_i \text{ satisfying }\\
    \frac{1}{\dt}(u^{h,(k)}_i-u^{h,n-1}_{i,snap},v^h)+(\sigma_i(u^{h,(k)}_i),\nabla v^h)=(f_i^n,v^h)
    +(-1)^{i}(g^{n,(k-1)},v^h)_{\Gz} \hspace{5pt}\forall v^h\in V_i^h,
\end{multline} or equivalently,
\begin{multline}\label{eqn:ModWFmatrix}
    \text{find } \bu_i^{(k)}\in \R^{N_i} \text{ satisfying }\\
    \frac{1}{\dt}M_i(\bu^{(k)}_i-\bu^{n-1}_{i,snap})+(\nu_i K_i-A_i)\bu^{(k)}_i=\bar{f}_i
    +(-1)^{i}M_{\Gz,i}\bar{g}^{n,(k-1)} 
\end{multline}
}where $M_i$, $K_i$, $A_i$, $M_{\Gz,i}$, and $\bar{f}_i$ are the same as previously defined, $\bar{u}^{n-1}_{i,snap}$ is the $(n-1)^{st}$ time step solution on $\Omega_i$ from the state snapshot, and $k$ is defined in Algorithm \ref{alg:ModkGDCont}. The modified gradient descent method for adjoint snapshot collection is given in Algorithm \ref{alg:ModkGDCont} and includes one or more steps of gradient descent which solve \eqref{eqn:ModWFmatrix} instead of \eqref{eqn:FEMwfmatrix}. \PAKe{Note that the solution $\bar{u}_i^{(k)}$ will not be used in the next optimization iteration at the same timestep, nor when solving this problem at the next time step. This effectively breaks the connection between solution of Algorithm \ref{eqn:ModWF}} over timesteps and allows for the computations to be made in parallel.

\begin{remark}
    Using MGDmRA, each time step is not dependent upon the previous time steps computed solutions. Therefore this algorithm is parallellizable in time. 
\end{remark}


\begin{algorithm}[H]
\caption{Modified gradient descent method for $m\ge1$ (MGDmRA)}\label{alg:ModkGDCont}
\begin{algorithmic}
    \State At a timestep $n$, initialize $\bar{g}^{n,(0)}$ 
    \For{$k=1,...,m$}\\
    \State Solve \EVHf{\eqref{eqn:ModWFmatrix}} using $\bar{g}^{n,(k-1)}$, $\bu^{n-1}_{1,snap}$, and $\bu^{n-1}_{2,snap}$ for $\bu^{(k)}_1$ and $\bu^{(k)}_2$\\
    \State Solve \eqref{eqn:FEMadjointmatrix} using $\bu^{(k)}_1$ and $\bu^{(k)}_2$ for $\bmu_1^{(k)}$ and $\bmu_2^{(k)}$\\
    \State $\bar{g}^{n,(k)}= (1-\alpha\delta)\bar{g}^{n,(k-1)} -\alpha(\bmu^{(k)}_1-\bmu^{(k)}_2)|_{\Gz}$\\
    \State Store $\bar{\mu}_1^{(k)}$ and $\bar{\mu}_2^{(k)}$ in their respective adjoint snapshot matrices
    \EndFor

\end{algorithmic}

\end{algorithm}

\section{Numerical Results}\label{sec:numericalresults}

\begin{figure}[H]
    \centering
     \includegraphics[scale=.3]{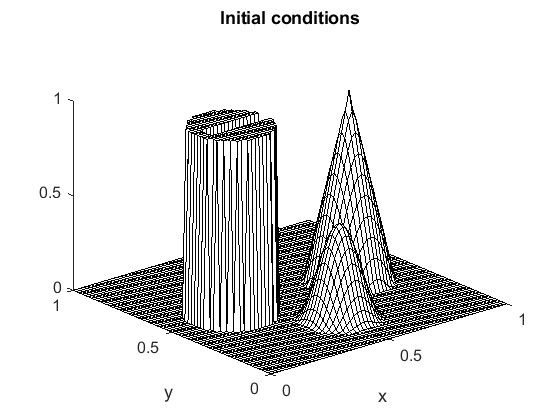}
    \includegraphics[scale=1.25]{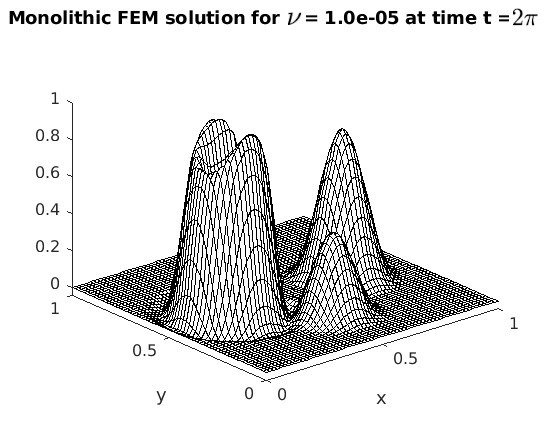}
    \caption{\EVHR{ Left: initial conditions and right: monolithic FEM solution at the final time step.}}
    \label{fig10_2}
\end{figure}

\PBB{To demonstrate our OBC approach with the MGD$m$RA snapshot collection procedure, we consider the model problem \eqref{eqn:ad-fully-coupled-problem} and the solid body rotation test from \cite{Leveque_96_SINUM}. In this test $\mathbf{a}_i(x,t):=(0.5-y,x-0.5)$ is a rotational velocity field and the initial condition comprises a notched cylinder, a cone, and a Gaussian hill; see Figure \ref{fig10_2}.  We take $\Omega$ to be the unit square and define the non-overlapping subdomains as $\Omega_1=(0,0.5)\times (0,1)$ and $\Omega_2=(0.5,1)\times (0,1)$. The interface $\Gamma_0$ is thus the line segment $\{(x,y) | x=0.5, 0\le y\le 1\}$.} \PAKr{We let $\nu_i=10^{-5}$ for this test, however, in later experiments we will also explore $\nu_i=10^{-3}$ and this will be noted.}

\PBB{We set the final simulation time to correspond to one full rotation of the initial condition, i.e.,} $T=2\pi$, and discretize in time using backward Euler with $\dt= 1.122398\cdot 10^{-3}$. 
\PBB{We use this time step for accuracy rather than stability reasons, as the backward Euler scheme is unconditionally stable.} 
\PBB{To discretize the model problem in space we use standard Lagrangian $\mathbb{Q}^1$ elements defined on a $64\times 64$ partition of $\Omega$ into quadrilateral finite elements with 4225 DoFs. The finite element mesh is interface conforming, i.e., the interface $\Gamma_0=\{(x,y) | x=0.5, 0\le y\le 1\}$ is one of the vertical grid lines. In so doing, we ensure that the restriction of the \PAKr{mesh} to each subdomain defines a conforming finite element partition of $\Omega_i$ having 2145 DoFs.}

\PBB{We define a reference solution by solving a monolithic formulation of \eqref{eqn:ad-fully-coupled-problem} using the same $64\times 64$ mesh and $\mathbb{Q}^1$ elements. Thanks to the interface-conforming mesh choice, the monolithic solution can be easily split into parts corresponding to each subdomain.}

For gradient descent, we use $\delta=10^{-16}$ and $\alpha=2$, which is reduced as needed within the algorithm if \PAKe{a step size is too large in a descent direction}.
\PBB{We recall the objective function \eqref{eqn:functional},  which measures the state mismatch on the interface and penalizes the size of the control. We use the value of this function to define a stopping criterion for the optimizer.} 
\PAKe{We set the value of $\delta$} \PBB{in \eqref{eqn:functional} to be less than } the \PAKe{convergence tolerance, e.g., with a tolerance of $1e-14$ we set $\delta=1e-16$ and expect to be able to find a control with $\|u_c-u_m\|\approx O(10^{-\EVHf{7}})$ at an optimal solution.} \PBB{Here $u_c$ denotes the OBC solution and $u_m$ is the monolithic solution.}

We begin by solving the FOM-FOM coupled problem and comparing it to the monolithic solution. 

\begin{figure}[H]
    \centering
    \includegraphics[scale=0.5]{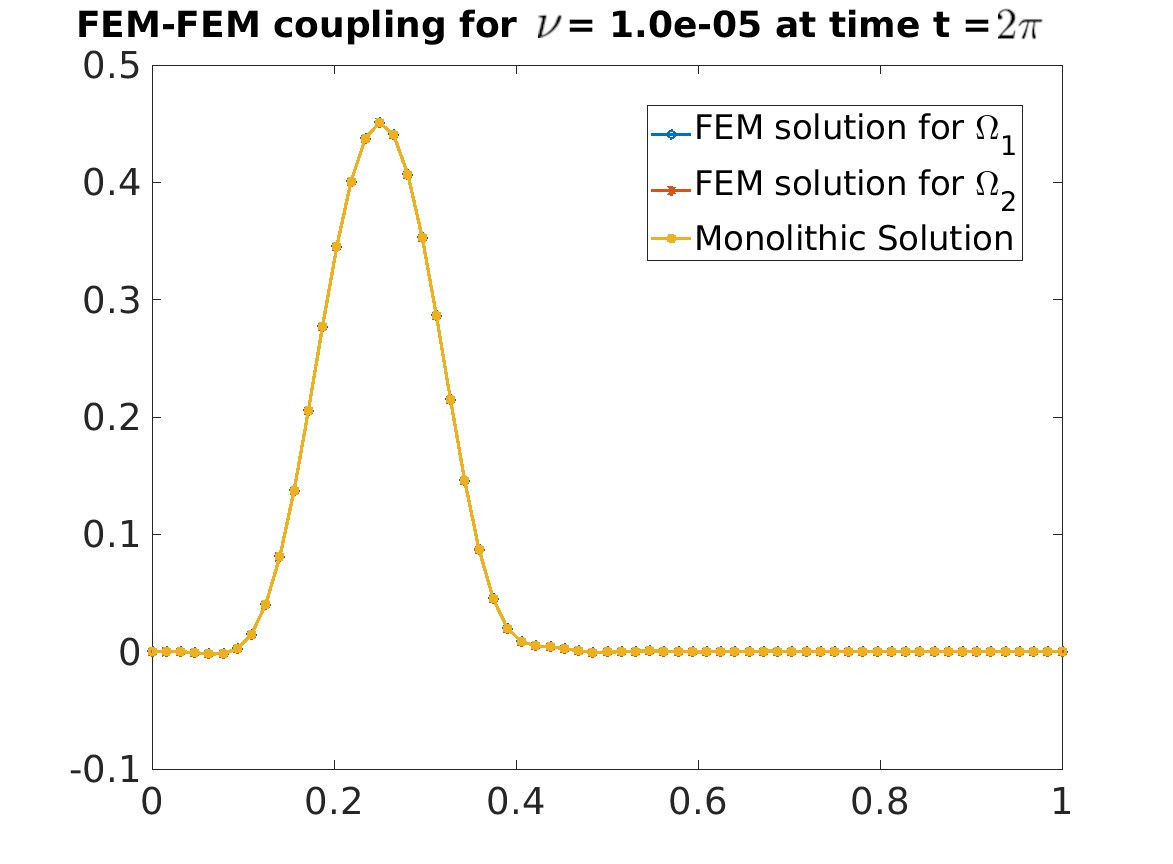}
    \includegraphics[scale=.55]{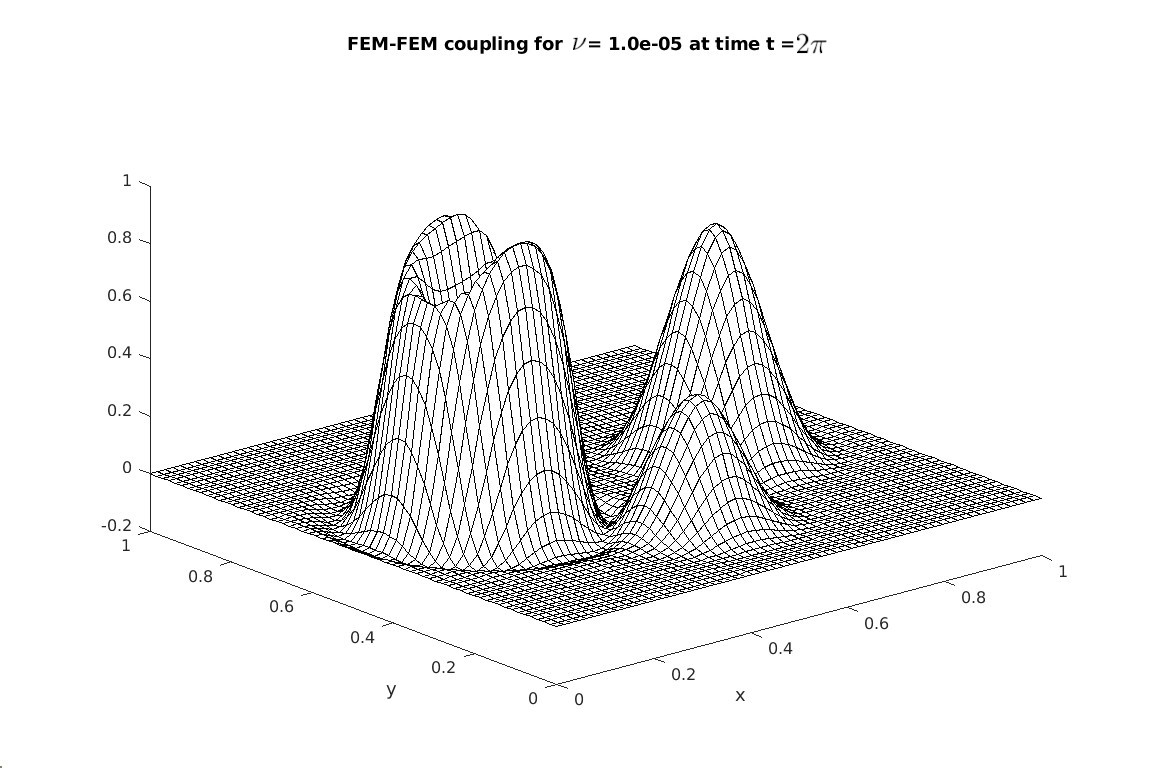}
    \caption{$u_m$ and $u_c$ on the interface and $u_c$ on the entire domain.}
    \label{fig:fig10}
\end{figure}

For the FOM-FOM coupled problem \PBB{we find that}
$$
    \frac{\|u_c - u_m\|_{L^2} }{ \|u_m\|_{L^2} } = \EVHd{7.8\cdot 10^{-8}}
    \quad\mbox{and}\quad
    \frac{\|u_c - u_m\|_{H^1} }{ \|u_m\|_{H^1} } = \EVHd{2.9\cdot 10^{-7}}\,,
$$
\PBB{respectively}.
This is the expected accuracy for $u_c$, as explained previously, and it can be made more accurate by reducing \EVHf{$\delta$ and} \PAKe{the convergence tolerance}. \EVHR{Furthermore, we see in Figure \ref{fig:fig10} that the shape of the coupled solution matches the monolithic solution on the interface (left) and the entire domain (right).}

\begin{figure}[H]
    \centering
    \includegraphics[scale=0.35,clip,trim=3cm 8cm 4cm 8cm]{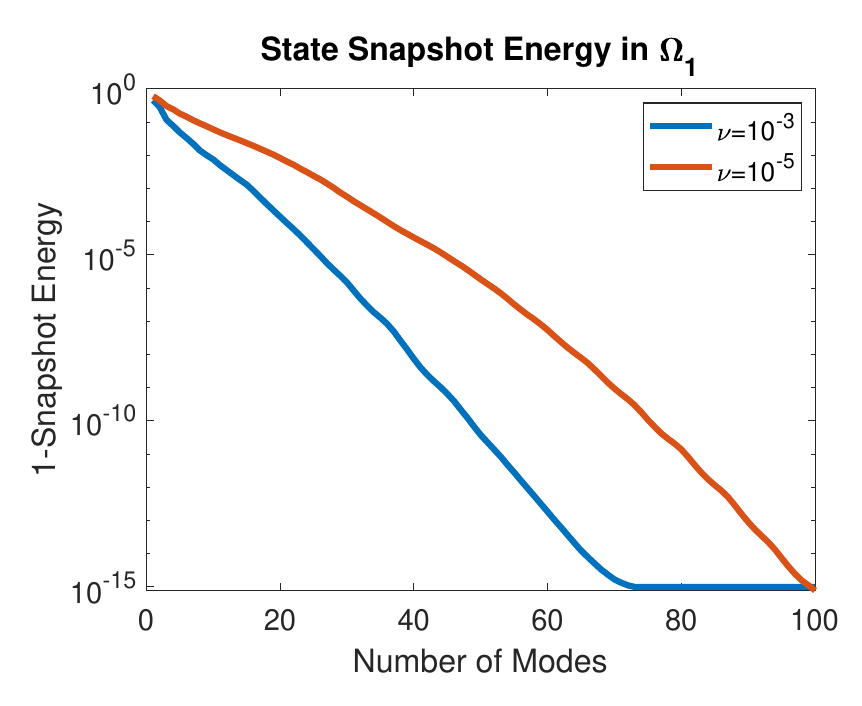}
    \includegraphics[scale=0.35,clip,trim=3cm 8cm 4cm 8cm]{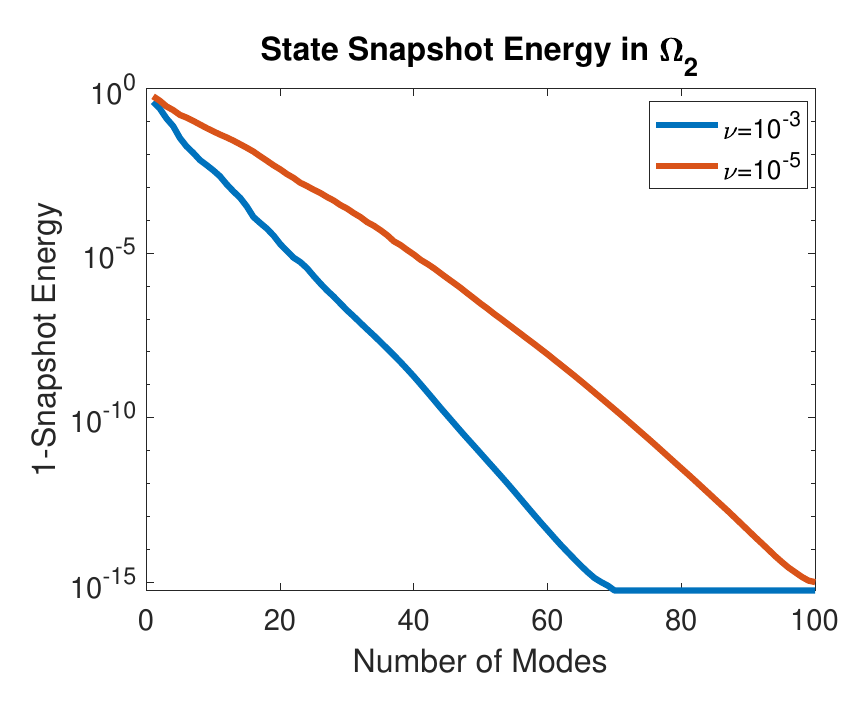}
    \caption{Comparison of snapshot energy for \PAKr{state solutions} in $\Omega_1$ (left) and $\Omega_2$ (right), for $\nu=\{1e-3,1e-5\}$, as a function of the number of modes retained.}
    \label{fig:StateSingValues}
\end{figure}

\begin{figure}[H]
    \centering
    \includegraphics[scale=0.35,clip,trim=3cm 8cm 4cm 8cm]{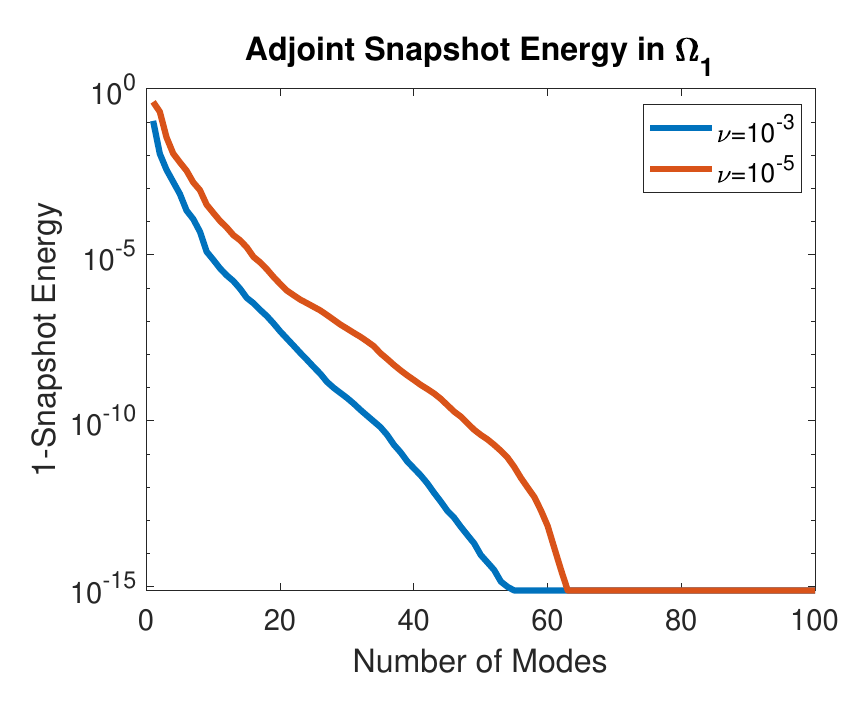}
    \includegraphics[scale=0.35,clip,trim=3cm 8cm 4cm 8cm]{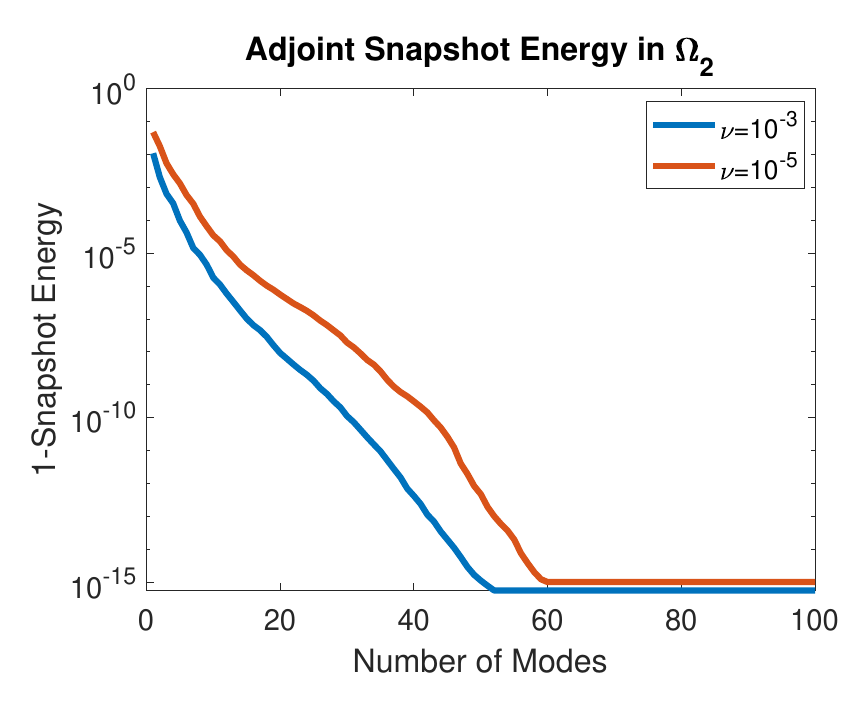}
    \caption{Comparison of snapshot energy for \PAKr{adjoint solutions} in $\Omega_1$ (left) and $\Omega_2$ (right), for values of $\nu$ as a function of the number of modes retained.}
    \label{fig:GDAdjSingValues}
\end{figure}

We collect snapshots of the state solutions and plot \PAKr{a function of} the snapshot energy in Figure \ref{fig:StateSingValues} for  $\nu=\{1e-3,1e-5\}$. We do the same for snapshots energies of the FOM-FOM adjoint system solutions in Figure \ref{fig:GDAdjSingValues}. The snapshot energies in each figure rapidly increase with number of modes retained, indicating that both the state and adjoint systems are good candidates for projection-based reduced order modeling.


Before testing the  MGD$m$RA snapshot collection approach and comparing it with the use of the state ROM basis for the adjoint, let us establish a reference \PAKr{solution by using ROM states for the reduced primal problem and using a non-truncated reduced order model for the adjoint system of equations with the reduced basis generated from snapshots of the adjoint solutions, which we will refer to with the initialism ntRA (non-truncated reduced-order model for the adjoint). Similarly, when using a non-truncated reduced-order model for the state system of equations with reduced basis generated from snapshots of the state solutions, we refer to it with the initialism ntRS.}
\PAKr{Using the non-truncated reduced-order adjoint (ntRA) along with a reduced state system allows} us to isolate the effects prompted by the use of a reduced basis \PAKr{for the} adjoint.
\PBB{Results shown in Table \ref{tab:RSFA} confirm that when the optimizer uses the \PAKr{non-truncated reduced-order adjoint}, the coupled ROM-ROM solution achieves the same accuracy as the coupled FOM-FOM solution, provided we use enough ($>50$) modes in the reduced basis \PAKr{for the primal problem}.} 



\begin{remark}
    In the numerical results that follow, we use $\ast$ to denote when the gradient descent algorithm exceeds 10,000 iterations of gradient descent in a single timestep. This is many more iterations of gradient descent than one would normally execute in practice. In these cases, we provide the error that the algorithm is able to attain and denote this behavior by $\ast$. Note that this behavior implies that there is stagnation in the gradient descent algorithm for these modal amounts due to the ROM-ROM model lacking sufficient accuracy to continue converging.
\end{remark}
\begin{remark}
    \EVHR{We also tested that the approach passes a patch test with a manufactured solution that is linear in both time and space, using $\delta=0$ and convergence tolerance $10^{-27}$. 
    Solution error for the ROM-ROM coupled problem is
    \begin{align*}
        \frac{\|u_c-u_m\|_{L^2}}{\|u_m\|_{L^2}}=2.006\cdot 10^{-14}\\
        \frac{\|u_c-u_m\|_{H^1}}{\|u_m\|_{H^1}}=1.120\cdot 10^{-12}\\
    \end{align*}
    when 500 and 250 modes are retained for the state and adjoint reduced bases, respectively.}
\end{remark}

\begin{table}[H]
    \centering
    \begin{tabular}{|c|c|c|}
    \hline
    & \multicolumn{2}{c|}{RS-ntRA}\\
    \hline
    Modes & Error & Avg. Iters.\\
    \hline
    50& $10^{-4}$& $\ast$\\
    100& $10^{-7}$& \EVHd{50.1}\\
    250& $10^{-7}$& \EVHd{36.9}\\
    500& $10^{-7}$& \EVHd{21.4}\\
    1000& $10^{-7}$& \EVHd{9.8}\\
    1500& $10^{-8}$& \EVHd{4.4}\\
    1600& $10^{-8}$& \EVHd{4.4}\\
    1700& $10^{-8}$& \EVHd{4.4}\\
    1800& $10^{-8}$& \EVHd{4.4}\\
    2016& $10^{-8}$& \EVHd{3}\\
    \hline
    \end{tabular}
    \caption{Error $\EVHd{\frac{\|u_c-u_m\|}{\|u_m\|}}$ and \PAKr{average of} iteration counts \PAKr{over timesteps} for ROM-ROM coupled problem with $\delta=10^{-16}$, $\nu=10^{-5}$, and convergence tolerence $10^{-14}$ for reduced space + non-truncated reduced-order adjoint (RS-ntRA).}
    \label{tab:RSFA}
\end{table}

\begin{table}[H]
    \centering
    \EVHd{\begin{tabular}{|c|c|c|}
    \hline
     & \multicolumn{2}{c|}{RS}\\ 
    \hline
    Modes & $\mathcal{E}(u_1,\Psi_{u,1})$ & $\mathcal{E}(u_2,\Psi_{u,2})$\\
    \hline 
        50 & $10^{-3}$ & $10^{-3}$\\
        100 & $10^{-8}$ & $10^{-7}$\\
        250 & $10^{-9}$ & $10^{-8}$\\
        500 & $10^{-9}$ & $10^{-8}$\\
        1000 & $10^{-9}$ & $10^{-8}$\\
        1500 & $10^{-9}$ & $10^{-8}$ \\
        2016 & $10^{-15}$ & $10^{-15}$\\
         \hline
    \end{tabular}}
    \caption{Projection error of the state solutions in $\Omega_1$ and $\Omega_2$ onto the reduced basis generated from state solution snapshots with $\delta=10^{-16}$, $\nu=10^{-5}$, and convergence tolerance $10^{-14}$.}\label{tab:state-projection-of-state}
    \end{table}

\subsection{State ROM Basis in the Adjoint}
\PBB{We first present results for the ROM-ROM optimization-based coupling when the gradient descent algorithm uses a ROM basis for the adjoint defined from the already available state solution snapshots.} To see the effectiveness of the state ROM basis in the adjoint, we first run the ROM-ROM coupled problem with a non-truncated reduced-order state and the state ROM basis in the adjoint (ntRS-SRA). 

We see in the last \PAKe{column} of Table \ref{tab:statetable} that the state ROM basis for the adjoint \PAKe{requires upwards of 500 modes to achieve accurate results.} This behavior is also reflected in the ROM-ROM coupled problem results with a ROM state and ROM state basis for the adjoint (RS-SRA) in Table \ref{tab:statetable}. However, as shown in the ROM state and \PAKr{non-truncated reduced-order adjoint in} Table \ref{tab:RSFA}, we see that the ROM state is not the cause of the deterioration in RS-SRA, since RS-ntRA provides accurate results for as few as 100 modes.

\PAKr{Table \ref{tab:state-projection-of-state} confirms that the reduced basis generated from state snapshots has low projection error for state solutions, as expected, while Table \ref{tab:StateProj} demonstrates that the same reduced bases have high projection error for adjoint solutions. Note that keeping any fewer than every possible mode for the reduced basis for projection of adjoint solutions in Table \ref{tab:StateProj} gives a large loss in accuracy with projection error $O(10^{-3})$ or greater. This indicates that state solution snapshots are not appropriate for use in generating a reduced basis for the adjoint system. }

Considering all modes being required for an accurate projection of the adjoint systems in Table \ref{tab:StateProj} and the far fewer iterations required for RS-ntRA to converge compared to ntRS-SRA and RS-SRA, it \PAKe{is} clear \PAKe{that using the state solutions snapshots to form a reduced basis for the adjoint systems is not appropriate.}


\begin{table}[H]
    \centering
    \begin{tabular}{|c|c|c|c|c|}
    \hline
    & \multicolumn{2}{c|}{RS-SRA} & \multicolumn{2}{c|}{ntRS-SRA}\\
    \hline
         Modes & Error & Avg. Iters & Error & Avg. Iters. \\
         \hline
         50& $10^{-3}$& $\ast$& $10^{-4}$& $\ast$\\
         100& $10^{-3}$& $\ast$& $10^{-4}$& $\ast$\\
         250& $10^{-5}$& $\ast$& $10^{-4}$& $\ast$\\
         500& $10^{-6}$& $\ast$& $10^{-5}$& $\ast$\\ 
         1000& $10^{-7}$& $\ast$& $10^{-6}$& $\ast$\\
         1500& $10^{-7}$& \EVHd{506.7}& $10^{-8}$& \EVHd{139.8}\\
         1600& $10^{-8}$& \EVHd{68.6}& $10^{-8}$& \EVHd{48.2}\\
         1700& $10^{-8}$& \EVHd{21.1}& $10^{-8}$& \EVHd{11.9}\\
         1800& $10^{-8}$& \EVHd{14.8}& $10^{-8}$& \EVHd{9}\\
         2016& $10^{-8}$& \EVHd{3}& $10^{-8}$& \EVHd{3}\\
         \hline
    \end{tabular}
    
    \caption{Error $\EVHd{\frac{\|u_c-u_m\|}{\|u_m\|}}$ and \PAKr{average of} iteration counts \PAKr{over timesteps} for ROM-ROM coupled problem, with $\delta=10^{-16}$, \EVHd{tolerance $10^{-14}$}, and $\nu=10^{-5}$. \PAKp{Reduced state + state snapshot reduced adjoint (RS-SRA) and non-truncated reduced-order state + state snapshot reduced adjoint (ntRS-SRA).}}
    \label{tab:statetable}
\end{table}

\begin{table}[H]
    \centering
    \begin{tabular}{|c|c|c|}
    \hline
     & \multicolumn{2}{c|}{SRA}\\ 
    \hline
    Modes & $\mathcal{E}(\mu_1,\Psi_{u,1})$ & $\mathcal{E}(\mu_2,\Psi_{u,2})$\\
    \hline 
        50 & $10^{-1}$ & $10^{-1}$\\
        2000 & $10^{-1}$ & $10^{-1}$\\
        2001 & $10^{-2}$ & $10^{-2}$\\
        2014 & $10^{-2}$ & $10^{-2}$\\
        2015 & $10^{-3}$ & $10^{-3}$ \\
        2016 & $10^{-15}$ & $10^{-15}$\\
         \hline
    \end{tabular}
    \caption{Projection error of the adjoint solutions in $\Omega_1$ and $\Omega_2$ for SRA with $\delta=10^{-16}$, $\nu=10^{-5}$, and convergence tolerence $10^{-14}$. Note that two numbers separated by a dash represents a range of quantities that varies over timesteps and gradient descent iterations.}
    \label{tab:StateProj}
    \end{table}

\subsection{Modified Gradient Descent ROM basis for the Adjoint}
\PBB{We now proceed to examine the new MGD$m$RA snapshot collection approach. Our results are summarized in Table \ref{tab:MGDtable}. From the data in this table we see that, \EVHR{for more than 100 modes,} the} MGD$m$RA ROM basis constructed by collecting \PBB{$m=1$} \PAKe{adjoint} snapshot per time step (MGD1RA) provides results that are as accurate as the FOM\PAKr{-FOM coupled problem}. 

We combine MGD$m$RA ROM adjoint with a ROM state and, as expected, the iteration count increases as the modes decrease. 
Finally, we check the projection error of the basis for the adjoint in Table \ref{tab:MGDProj} and see that we get $O(10^{-15})$ projection error for more than 100 modes with some variation between gradient descent steps and timesteps. Below 100 modes, we see worse projection errors which matches the results in the previous table.

We also check the projection errors of the MGD$m$RA ROM basis constructed by collecting \PAKe{one snapshots per timestep (MGD1RA) and the MGD$m$RA ROM basis constructed by collecting two snapshots per timestep (MGD2RA)}. As shown in Table \ref{tab:MGDProj}, the projection errors do not significantly change by keeping more adjoint snapshots. Instead when a range of projection errors occurs, the distribution of the projection errors changes. This has almost no affect on the iteration count and accuracy of the ROM-ROM coupled problem.


\begin{table}[H]
    \centering
    \begin{tabular}{|c|c|c|c|c|}
    \hline
    & \multicolumn{2}{c|}{RS-MGD1RA} & \multicolumn{2}{c|}{ntRS-MGD1RA}\\
    \hline
         Modes & Error & Avg. Iters. & Error & Avg. Iters.\\
         \hline
         50&  $10^{-4}$& $\ast$& $10^{-7}$& $\ast$\\
         100&  $10^{-5}$& \EVHe{2342.8}& $10^{-8}$& \EVHe{4.2}\\
         250& $10^{-6}$& \EVHe{401.5}& $10^{-8}$& \EVHe{4}\\
         500&  $10^{-7}$& \EVHe{102.8}& $10^{-8}$& \EVHe{3.7}\\ 
         1000& $10^{-7}$& \EVHe{35.2}& $10^{-8}$& \EVHe{3.7}\\
         1500& $10^{-8}$& \EVHe{7.8}& $10^{-8}$& \EVHe{3.7}\\
         1600&  $10^{-8}$& \EVHe{6.6}& $10^{-8}$& \EVHe{3.7}\\
         1700&  $10^{-8}$& \EVHe{6.5}& $10^{-8}$& \EVHe{3.7}\\
         1800&  $10^{-8}$& \EVHe{6.5}& $10^{-8}$& \EVHe{3.7}\\
         2016&  $10^{-8}$& \EVHd{3}& $10^{-8}$& \EVHd{3}\\
         \hline
    \end{tabular}
    
    \caption{Error $\EVHd{\frac{\|u_c-u_m\|}{\|u_m\|}}$ and \PAKr{average of} iteration counts \PAKr{over timesteps} for ROM-ROM coupled problem, with $\delta=10^{-16}$, $\nu=10^{-5}$, and convergence tolerence $10^{-14}$. reduced state + modified gradient descent reduced adjoint (RS-MGD1RA) and non-truncated reduced-order state + modified gradient descent reduced adjoint (ntRS-MGD1RA).}
    \label{tab:MGDtable}
\end{table}

    \begin{table}[H]
    \centering
    {\small\begin{tabular}{|c|c|c|c|c|}
    \hline
    & \multicolumn{2}{c|}{MGD1RA} & \multicolumn{2}{c|}{MGD2RA}\\
    \hline
    Modes & $\mathcal{E}(\mu_1,\Psi_{\mu,1})$ & $\mathcal{E}(\mu_2,\Psi_{\mu,2})$& $\mathcal{E}(\mu_1,\Psi_{\mu,1})$ & $\mathcal{E}(\mu_2,\Psi_{\mu,2})$\\
    \hline 
        50 & $\EVHe{10^{-5}-10^{-2}}$ & $\EVHe{10^{-6}-10^{-3}}$& $\EVHe{10^{-6}-10^{-4}}$ & $\EVHe{10^{-6}-10^{-4}}$\\
        100 & $\EVHe{10^{-15}}-\EVHe{10^{-12}}$ & $\EVHe{10^{-15}}-\EVHe{10^{-13}}$& $\EVHe{10^{-15}}-\EVHe{10^{-13}}$ & $\EVHe{10^{-15}}-\EVHe{10^{-14}}$\\
        500 & $\EVHe{10^{-15}}-\EVHe{10^{-12}}$ & $\EVHe{10^{-15}}-\EVHe{10^{-13}}$& $\EVHe{10^{-15}}-\EVHe{10^{-13}}$ & $\EVHe{10^{-15}}-\EVHe{10^{-14}}$\\
        1500 & $\EVHe{10^{-15}}-\EVHe{10^{-12}}$ & $\EVHe{10^{-15}}-\EVHe{10^{-13}}$& $\EVHe{10^{-15}}-\EVHe{10^{-12}}$ & $\EVHe{10^{-15}}-\EVHe{10^{-13}}$\\
        2016 & $10^{-15}$ & $10^{-15}$& $10^{-15}$ & $10^{-15}$\\
         \hline
    \end{tabular}}
    \caption{Projection error of the adjoint solutions in $\Omega_1$ and $\Omega_2$ for \PAKe{MGD1RA and MGD2RA} with $\delta=10^{-16}$, $\nu=10^{-5}$, and convergence tolerence $10^{-14}$. Note that two numbers separated by a dash represents a range of quantities that varies over timesteps and gradient descent iterations.}
    \label{tab:MGDProj}
    \end{table}

\subsubsection{Different Modal Numbers in State ROM and Adjoint ROM}

Because \EVHd{MGD$m$RA} adjoint ROM does not need as many modes as the state ROM, we investigate using a different number of modes in \PAKr{the reduced-order state (RS)} and MGD$m$RA adjoint ROM. As seen in Table \ref{tab:RSFA}, using 100 or more modes in the \PAKr{RS} results in the same accuracy as the state FOM. Therefore, in the following tables we will keep between 50 and 250 modes in the \PAKr{RS}. Similarly, as seen in Table \ref{tab:MGDtable}, keeping 100 or more modes in the MGD$m$RA adjoint ROM results in the same accuracy as the non-truncated reduced-order adjoint. As a result, we will not keep more than 100 modes in the MGD$m$RA adjoint ROM in the following tables.

We begin using the same tolerance, $\EVHd{10^{-14}}$, as the previous tables. In Table \ref{tab:stateMGDtable1}, we see that when keeping 25 MGD1RA adjoint ROM modes or 50 state modes we do not attain convergence and do not attain accuracy near $10^{-8}$. Once we keep 50 or more MGD1RA adjoint ROM modes and 100 or more state modes, we do attain convergence. Note that keeping 50 state ROM modes never achieves convergence or accuracy near $10^{-8}$ for any MGD1RA adjoint ROM modes. This is due to the state ROM needing more modes and is not due to the adjoint ROM needing more modes which is clearly illustrated in the Table \ref{tab:stateMGDtable1}. 
\PAKe{Increasing the number of MGD1RA adjoint ROM modes and the state ROM modes concurrently deacreases the number of iterations needed, and Tables \ref{tab:stateMGDtable1}-\ref{tab:stateMGDtable3} seem to suggest there is a balance to be struck, i.e., it is not always beneficial to increase the MGD1RA adjoint ROM modes while keeping the state ROM modes fixed nor vice-versa.}

When using relatively few modes, it is common that a ROM is unable to attain the same accuracy as \PAKr{a} FOM. Therefore, we provide a table using \EVHd{$\delta=10^{-14}$} with the tolerance $10^{-12}$. \PAKr{I}n Table \ref{tab:stateMGDtable2} we see the same results as Table \ref{tab:stateMGDtable1}, but with lower iteration counts. 
Again, in Table \ref{tab:stateMGDtable3}, we lower the tolerance to $10^{-8}$ and \EVHd{simultaneously lower $\delta$ to $10^{-10}$}. 


\begin{table}[H]
    \centering
    \begin{tabular}{|c|c|c|c|c|c|c|}
    \hline
         RS Modes& \multicolumn{2}{c|}{50}& \multicolumn{2}{c|}{100}& \multicolumn{2}{c|}{250}\\
         \hline
         MGD1RA Modes& Avg. Iters.& Error& Avg. Iters.& Error& Avg. Iters.& Error\\
         \hline
         \EVHd{25}&$\ast$ & $10^{-4}$ & \EVHe{4498.8}& $10^{-6}$ & $\ast$& $10^{-4}$ \\
         50& $\ast$& $10^{-4}$& \EVHe{761.5}& $10^{-6}$& \EVHe{2119.8}& $10^{-6}$\\
         100& $\ast$& $10^{-4}$& \EVHe{2342.8}& $10^{-5}$& \EVHe{536.9}& $10^{-6}$\\
         \hline
    \end{tabular}
    
    \caption{Error $\EVHd{\frac{\|u_c-u_m\|}{\|u_m\|}}$ and \PAKr{average of }iteration counts \PAKr{over timesteps} for ROM-ROM coupled problem, with $\delta=10^{-16}$, convergence tolerance $10^{-14}$, and $\nu=10^{-5}$ for varying modes in reduced state + modified gradient descent reduced adjoint (RS-MGD1RA).}
    \label{tab:stateMGDtable1}
\end{table}

\begin{table}[H]
    \centering
    \begin{tabular}{|c|c|c|c|c|c|c|c|c|}
    \hline
         RS Modes& \multicolumn{2}{c|}{50}& \multicolumn{2}{c|}{100}& \multicolumn{2}{c|}{250}\\
         \hline
         MGD1RA Modes& Avg. Iters.& Error& Avg. Iters.& Error& Avg. Iters.& Error\\
         \hline
         \EVHd{25}& $\ast$ & $10^{-4}$& \EVHe{349}& $10^{-5}$& \EVHe{4246.4}& $10^{-6}$\\
         50& $\ast$& $10^{-4}$& \EVHe{170.7}& $10^{-5}$& \EVHe{312.4}& $10^{-6}$\\
         100& $\ast$& $10^{-4}$& \EVHe{150.4}& $10^{-5}$& \EVHe{129.5}& $10^{-5}$\\
         \hline
    \end{tabular}
    
    \caption{Error $\EVHd{\frac{\|u_c-u_m\|}{\|u_m\|}}$ and \PAKr{average of }iteration counts \PAKr{over timesteps} for ROM-ROM coupled problem, with $\delta=10^{-14}$, $\nu=10^{-5}$, and convergence tolerance $10^{-12}$ for varying modes in reduced state + modified gradient descent reduced adjoint (RS-MGD1RA).}
    \label{tab:stateMGDtable2}
\end{table}

\begin{table}[H]
    \centering
    \begin{tabular}{|c|c|c|c|c|c|c|c|c|}
    \hline
         RS Modes& \multicolumn{2}{c|}{50}& \multicolumn{2}{c|}{100}& \multicolumn{2}{c|}{250}\\
         \hline
         MGD1RA Modes& Avg. Iters.& Error& Avg. Iters.& Error& Avg. Iters.& Error\\
         \hline
         \EVHd{25}& \EVHe{15.2}&$10^{-3}$ &\EVHe{9.3} &$10^{-4}$& \EVHe{15.6} & $10^{-4}$\\
         50& \EVHe{18.9}& $10^{-3}$& \EVHe{11.4}& $10^{-4}$& \EVHe{8.8}& $10^{-4}$\\
         100& \EVHe{12.7}& $10^{-3}$& \EVHe{8.4}& $10^{-4}$& \EVHe{6.2}& $10^{-4}$\\
         \hline
    \end{tabular}
    
    \caption{Error $\EVHd{\frac{\|u_c-u_m\|}{\|u_m\|}}$ and \PAKr{average of }iteration counts \PAKr{over timesteps} for ROM-ROM coupled problem, with $\delta=10^{-10}$ and $\nu=10^{-5}$ and convergence tolerance $10^{-8}$ for varying modes in reduced state + modified gradient descent reduced adjoint (RS-MGD1RA).}
    \label{tab:stateMGDtable3}
\end{table}

\subsubsection{Comparison to Gradient Descent Adjoint Snapshots}\label{GDMGDcomparison}
The usual snapshot collection method for adjoints is to solve the FOM-FOM coupled problem and store all computed adjoints for all time steps; this is the gradient descent for reduced adjoint (GDRA) snapshot method. Because MGD$m$RA does not compute or \PAKe{store the adjoint snapshots from all iterations at all timesteps} for the coupled problem, it is important to compare MGD$m$RA to GDRA for any \PAKe{e}ffect caused by this reduction or by the variation in the adjoint problem.


In Table \ref{tab:GDProj}, we see that projection errors for GDRA ROM have the same order of errors, or range of errors, for each amount of modes as MGD1RA. It is important to remark, though, that the projection errors at each gradient descent step for each time step are not exactly the same for MGD1RA and GDRA. In fact, the order of the projection errors may be different. However, the range of projection errors are the same.

    \begin{table}[H]
    \centering
      \begin{tabular}{|c|c|c|c|c|}
    \hline
     &\multicolumn{2}{c|}{GDRA}&\multicolumn{2}{c|}{MGD1RA}\\
    \hline
    Modes & $\mathcal{E}(\mu_1,\Psi_{\mu,1})$ & $\mathcal{E}(\mu_2,\Psi_{\mu,2})$& $\mathcal{E}(\mu_1,\Psi_{\mu,1})$ & $\mathcal{E}(\mu_2,\Psi_{\mu,2})$\\
    \hline 
        50 &  $\EVHe{10^{-4}-10^{-3}}$ & $\EVHe{10^{-6}-10^{-5}}$& $\EVHe{10^{-4}-10^{-3}}$ & $\EVHe{10^{-6}-10^{-5}}$\\
        100 & $\EVHe{10^{-15}-10^{-14}}$& $\EVHe{10^{-15}}$& $\EVHe{10^{-15}-10^{-14}}$& $\EVHe{10^{-15}}$\\
        500 & $\EVHe{10^{-15}-10^{-14}}$& $\EVHe{10^{-15}}$& $\EVHe{10^{-15}-10^{-14}}$& $\EVHe{10^{-15}}$\\
        1500& $\EVHe{10^{-15}-10^{-14}}$& $\EVHe{10^{-15}}$ & $\EVHe{10^{-15}-10^{-14}}$& $\EVHe{10^{-15}}$ \\
        2016& $10^{-15}$& $10^{-15}$& $10^{-15}$& $10^{-15}$\\
         \hline
    \end{tabular}
    \caption{Projection error of the adjoint solutions in $\Omega_1$ and $\Omega_2$ for GDRA \EVHe{and MGD1RA} with $\delta=\EVHd{10^{-14}}$, $\nu=10^{-5}$, and convergence tolerance $10^{-12}$. Note that two numbers separated by a dash represents a range of quantities that varies over timesteps and \EVHf{gradient descent} \PAKe{iterations}.}
    \label{tab:GDProj}
\end{table}

We next compare the MGD1RA ROM-ROM coupled problem and GDRA ROM-ROM coupled problem for $\EVHd{\delta=10^{-14}}$ and tolerance $10^{-12}$. Note that the FOM-FOM coupled problem to be solved as part of GDRA has an average of \EVHe{1.98} iterations per time step. Therefore GDRA stores approximately \EVHe{1.98} adjoint solutions per times step in the snapshot. However, MGD1RA stores one adjoint solution per time step. So there is a noticeable reduction in the size of the MGD1RA snapshot matrix. We compare the accuracy of the final solution given by $\frac{\|u_c-u_m\|}{\|u_m\|}$ and the average iteration per time step. Recall that with tolerance $10^{-12}$ we expect accuracy around $10^{-6}$. In Table \ref{tab:GDMGDtable} we see that MGD1RA ROM and GDRA ROM attain the same accuracy for each amount of modes. Furthermore, in Table \ref{tab:GDMGDtable} we see that there is some variation in the average number of iterations per time step, but it is not significant over the various modal amounts considered.



\begin{table}[H]
    \centering
    \begin{tabular}{|c|c|c|c|c|c|}
    \hline
         RS Modes& \multicolumn{2}{c|}{100}& \multicolumn{2}{c|}{250}\\
         \hline
         MGD1RA Modes& Avg. Iters.& Error& Avg. Iters.& Error\\
         \hline
         50& \EVHe{170.7}& $10^{-5}$& \EVHe{312.4}& $10^{-6}$\\
         100& \EVHe{150.4}& $10^{-5}$& \EVHe{129.5}& $10^{-5}$\\
         \hline
        GDRA Modes& & & & \\
         \hline
         50& \EVHe{170.7}& $ 10^{-5}$&\EVHe{371.2} & $10^{-6}$\\
         100& \EVHe{242.1}& $10^{-5}$& \EVHe{143.2}& $10^{-5}$\\
         \hline
    \end{tabular}

    \caption{Error $\EVHd{\frac{\|u_c-u_m\|}{\|u_m\|}}$ and \PAKr{average of }iteration counts \PAKr{over timesteps} for ROM-ROM coupled problem, with $\delta=\EVHd{10^{-14}}$, $\nu=10^{-5}$, and convergence tolerance $10^{-12}$ for varying modes in reduced state + gradient descent reduced adjoint (RS-GDRA) and reduced state + modified gradient descent reduced adjoint (RS-MGD1RA).}
    \label{tab:GDMGDtable}
\end{table}

\subsection{Timing study}

We perform a comparison of coupled ROM-ROM problems with different choices for the reduced basis for the adjoint problem. We compare using the reduced basis for adjoint from MGD1RA (one step of gradient descent) against a more traditional (but impractical) reduced basis for the adjoint based on all gradient descent steps (GDRA). 

With meshing spacing $h=\frac{1}{64}$ and $\Delta t=1.122398e-3$, we expect error in the FEM solution on the order of $1e-3$ (based on L2 FEM error on the order of $C(O(\Delta t)+O(h)^2$), so for this test we use a stopping criteria for the optimization algorithm of $1e-6$ (the square of $1e-3$). $\delta$ should be set sufficiently small so that $1e-6$ for the order of magnitude for the loss function is achievable, hence we set it to $1e-8$. 

We use the results from Tables \ref{tab:stateMGDtable1}, \ref{tab:stateMGDtable2}, and \ref{tab:stateMGDtable3} to inform our choice of the number of modes to keep for both the primal and adjoint reduced space systems. Based on Table \ref{tab:stateMGDtable3} particularly, we anticipate 100 or fewer modes being required for the reduced primal system, and 50 or fewer modes being required for the reduced adjoint system. While we think it is likely that 50 modes for each would provide a reasonably accurate solution, we will use 100 modes for the reduced primal system to provide as fair of a comparison as possible against the FOM-FOM optimization-coupled problem. 

Note that the FOM-FOM and ROM-ROM coupled problems solve the state and adjoint systems using an LU factorization that is computed one time, stored, and reused. For the ROM-ROM coupled problems this scales well to larger problems because the matrix dimensions can be reduced to the size that a direct method is favorable. However, for the FOM-FOM coupled problem, this does not scale well since there is no reduction in matrix dimensions as the problems size increases. This will ultimately lead to the FOM-FOM coupled problem requiring an iterative solver for the state and adjoint system solves. When the iterative solver for the systems is required, the FOM-FOM solver time will increase while in contrast the ROM-ROM will not require the iterative solver and therefore should scale better. 

We provide timings results in Table \ref{tab:timings}. For the FOM-FOM coupled problem, we get the solution at the final timestep in \EVHe{86} seconds with accuracy $O(10^{-3})$ as expected. The ROM-ROM coupled problem using MGD1RA and GDRA is solved in \EVHe{33} seconds and \EVHe{32} seconds, respectively, with $O(10^{-3})$ accuracy. Therefore, the ROM-ROM coupled problem is solved faster than the FOM-FOM coupled problem without any significant loss of accuracy. However, its important to note that with these parameters, the FOM-FOM coupled problem for the snapshot collection has an average of \EVHe{0.42} gradient descent iterations per time step. This means the only significant difference between MGD1RA and GDRA is the sequential nature of GDRA. 

For this reason, we consider the coupled problem with $\nu=10^{-3}$ for $\delta=\EVH{10^{-12}}$ and tolerance $\EVH{10^{-10}}$ where the FOM-FOM coupled problem is solved in an average of \EVHe{1.7} iterations per timestep. For these parameters, the FOM-FOM coupled problem is solved in \EVHe{131} seconds. We keep \EVHe{100} state modes and \EVHe{50} adjoint modes for the ROMs. The MGD1RA ROM-ROM coupled problem and GDRA ROM-ROM coupled problem are each solved in \EVHe{76} seconds. \EVH{Therefore, we again see that the MGD1RA ROM and GDRA ROM perform similarly and are faster than the FOM}. The parameter $\nu=10^{-5}$ makes the problem more \EVH{hyperbolic}, which is more difficult to capture with a small ROM basis but is easier to control. We also investigated $\nu=10^{-3}$, a more elliptic problem, which is \EVH{better suited for a small ROM basis,} but has a higher average iteration count.


\begin{table}[H]
    \centering
    \begin{tabular}{|c|c|c|c|c|c|}
    \hline
      \multicolumn{3}{|c|}{$\nu=10^{-5}$}& \multicolumn{3}{c|}{$\nu=10^{-3}$}\\
      \hline
      FOM& MGD1RA ROM & GDRA ROM& FOM& MGD1RA ROM & GDRA ROM\\
      \hline
       \EVHe{86} sec& \EVHe{33} sec& \EVHe{32} sec& \EVHe{131} sec & \EVHe{76} sec& \EVHe{76} sec\\
       \hline
    \end{tabular}
    \caption{Computational times for $\nu=10^{-5}$ and $\nu=10^{-3}$.}
    \label{tab:timings}
\end{table}

\subsection{Conclusions}
\PAKe{Adapting the optimization-based coupling} \PBB{(OBC) approach developed for full order models (FOMs) to the coupling or reduced order models (ROMs)} \PAKe{has the potential to decrease computation time to simulate coupled systems but also introduces a few challenges.} \PAKr{Solving the OBC PDE-constrained optimization problem by ``optimizing-then-reducing''} \PAKe{involves the solution of adjoint equations for which a reduced basis is not readily available.}

The most \PBB{straightforward} way of collecting adjoint snapshots is by solving the FOM-FOM coupled problem and storing all computed adjoints for all time steps. This \PAKe{snapshot strategy}, denoted GDRA, can result in large snapshot matrices and the snapshot collection algorithm \PAKe{is not obviously parallellizable in the sense that the primal solution at each timestep depends on the previous timestep solution, and therefore the FOM-FOM coupled system must be solved with a relatively tight tolerance to produce} \PBB{the adjoint snapshots}. \PAKe{This} requires a large amount of storage and computational time.

In comparison, \PAKe{the modified gradient descent for reduced basis for the adjoint (MGD$m$RA) snapshot collection technique produces a} matrix whose size is controllable by choosing how many \PAKe{iterations of gradient descent} to compute and store for each time step. 
Furthermore, the MGD$m$RA algorithm is parallellizable because each time step is not dependent upon the previous time step's solutions. \PAKe{This of course assumes that snapshots of the primal problem are available, as they are expected to be, for development of traditional projection-based reduced order models}. \PAKe{Both of these advantages lend themselves towards reduction of computation time for collecting snapshots from which to generate a reduced basis for the adjoint system}.

\PAKe{Numerical results indicate that MGD1RA is able to produce the same accuracy as the FOM for the adjoint problem using few modes, $N_{i,r} << N_i$. Accuracy for MGD2RA (keeping the adjoint snapshots for two iterations of gradient descent per timestep) shows there is little benefit over MGD1RA. Furthermore, the MGD$m$RA ROM does not have any detrimental affect on the ROM when compared to the much more expensive approach of performing gradient descent and keeping all adjoint snapshots from all iterations over all timesteps.} The accuracy of the MGD1RA ROM matches that of the GDRA ROM, and the average iterations per time step and run times for MGD1RA and gradient descent show no significant differences across all modes retained for the adjoint reduced basis. This similarity is also reflected in the computational time of MGD1RA and GDRA ROM, with both approaches outperforming the FOM-FOM coupled solution time. \PAKr{
In this paper, we used gradient descent in our algorithm for obtaining snapshots of adjoint solutions in a way that can be done parallel-in-time and without tight convergence criteria for the optimal control problem. However, the approach is sufficiently general so as to be applied with other optimization techniques, such as from the quasi-Newton family, that often converge in far fewer iterations. Also, in future work, we will consider the use of a reduced basis for the control space in addition to the subdomain problems, as this would likely yield further computational savings.}\\ 


\section*{Acknowledgments}
\PBB{This material is based upon work supported by the U.S. Department of Energy, Office of Science, Office of Advanced Scientific Computing Research, Mathematical Multifaceted Integrated Capability Centers (MMICCs) program, under Field Work Proposal 22-025291 (Multifaceted Mathematics for Predictive Digital Twins (M2dt)), Field Work Proposal 23-020467, and Computing and Information Sciences (CIS) investment area in the Laboratory Directed Research and Development program at Sandia National Laboratories.}

\PBB{This written work is authored by an employee of NTESS. The employee, not NTESS, owns the right, title and interest in and to the written work and is responsible for its contents. Any subjective views or opinions that might be expressed in the written work do not necessarily represent the views of the U.S. Government. The publisher acknowledges that the U.S. Government retains a non-exclusive, paid-up, irrevocable, world-wide license to publish or reproduce the published form of this written work or allow others to do so, for U.S. Government purposes. The DOE will provide public access to results of federally sponsored research in accordance with the DOE Public Access Plan.} SAND2024-11112O.

%
%

\bibliographystyle{siam}
\bibliography{ElizabethHawkins.bib}

\end{document}